\documentclass[twocolumn]{aastex61}

\newcommand{\Vmic}{\xi_{\rm t}}
\newcommand{\Vmac}{V_{\rm mac}}
\def\teff{T_{\rm eff}}
\def\logg{\log({\rm g})}

\newcommand{\mgi}{Mg\,{\textsc i}}

\newcommand{\md}{\langle {\rm 3D} \rangle}
\def\mgfe{[\hbox{Mg}/\hbox{Fe}]}
\newcommand{\fei}{Fe\,{\textsc i}}
\newcommand{\feii}{Fe\,{\textsc {ii}}}
\def\feh{\hbox{[Fe/H]}}

\received{}
\revised{}
\accepted{10 August, 2017}
\submitjournal{ApJ}

\shorttitle{Mg abundances in the Galactic disc and halo}
\shortauthors{Bergemann et al.}

\begin{document}


\title{Non-local thermodynamic equilibrium stellar spectroscopy with 1D and 3D models - II. Chemical properties of the Galactic metal-poor disc and the halo}

\correspondingauthor{Maria Bergemann}
\email{bergemann@mpia-hd.mpg.de}

\author{Maria Bergemann}
\affil{Max-Planck Institute for Astronomy, 69117, Heidelberg, Germany}

\author{Remo Collet}
\affil{Stellar Astrophysics center, Ny Munkegade 120, Aarhus University, DK--8000 Aarhus, Denmark}
\affil{Research School of Astronomy and Astrophysics, Australian National University, Canberra ACT 2601, Australia}

\author{Ralph Sch\"onrich}
\affil{Rudolf-Peierls center for Theoretical Physics, University of Oxford, 1 Keble Road, OX1 3NP, Oxford, United Kingdom}

\author{Rene Andrae}
\affil{Max-Planck Institute for Astronomy, 69117, Heidelberg, Germany}

\author{Mikhail Kovalev}
\affil{Max-Planck Institute for Astronomy, 69117, Heidelberg, Germany}

\author{Greg Ruchti}
\affil{Lund Observatory, Box 43, SE-221 00 Lund, Sweden}

\author{Camilla Juul Hansen}
\affil{Dark Cosmology Centre, Niels Bohr Institute, University of Copenhagen, Juliane Maries Vej 30, 2100 Copenhagen, Denmark}
\affil{Technische Universität Darmstadt, Schlossgartenstr. 2, Darmstadt D-64289, Germany}

\author{Zazralt Magic}
\affil{Niels Bohr Institute, University of Copenhagen, Juliane Maries Vej 30, DK--2100 Copenhagen, Denmark}
\affil{center for Star and Planet Formation, Natural History Museum of Denmark, {\O}ster Voldgade 5-7, DK--1350 Copenhagen, Denmark}

\begin{abstract}
From exploratory studies and theoretical expectations it is known that simplifying approximations in spectroscopic analysis (LTE, 1D) lead to systematic biases of stellar parameters and abundances. These biases depend strongly on surface gravity, temperature, and, in particular, for LTE vs. non-LTE (NLTE) on metallicity of the stars. Here we analyse the [Mg$/$Fe] and [Fe$/$H] plane of a sample of 326 stars, comparing LTE and NLTE results obtained using 1D hydrostatic models and averaged $\md$ models. We show that compared to the $\md$ NLTE benchmark, all other three methods display increasing biases towards lower metallicities, resulting in false trends of [Mg$/$Fe] against [Fe$/$H], which have profound implications for interpretations by chemical evolution models. In our best $\md$ NLTE model, the halo and disc stars show a clearer behaviour in the [Mg$/$Fe] $-$ [Fe$/$H] plane, from the knee in abundance space down to the lowest metallicities. Our sample has a large fraction of thick disc stars and this population extends down to at least [Fe$/$H] $\sim -1.6$ dex, further than previously proven. The thick disc stars display a constant [Mg$/$Fe] $\approx 0.3$ dex, with a small intrinsic dispersion in [Mg$/$Fe] that suggests that a fast SN Ia channel is not relevant for the disc formation. The halo stars reach higher [Mg$/$Fe] ratios and display a net trend of [Mg$/$Fe] at low metallicities, paired with a large dispersion in [Mg$/$Fe]. These indicate the diverse origin of halo stars from accreted low-mass systems to stochastic/inhomogeneous chemical evolution in the Galactic halo.%
\end{abstract}

\section{Introduction}
Chemical abundances in cool stars are perhaps the most important observational constraint to studies of chemical evolution of galaxies. In particular, in the Milky Way the abundance of the observed data, like that from the Gaia-ESO and APOGEE surveys, has lead to a major progress in our understanding of the formation of the Galaxy and its underlying stellar populations: the bulge, the disc, and the halo \citep[e.g.][]{mcwilliam1994,mcwilliam2008,nissen2010,casagrande2011,gonzalez2011,bensby2014,recioblanco2014,kordopatis2015,ness2016}. The observed chemical abundances are also essential tracers of chemical evolution and star formation of other galaxies \citep[e.g.][]{conroy2014,onodera2015,greene2015}.

Among the most critical discriminants of stellar populations in the Galaxy is the $\alpha$-enhancement, i.e. the relative abundance of $\alpha$-group elements to iron. In this respect, magnesium plays a key role as the classical $\alpha$-chain element, produced in hydrostatic burning in massive stars that end their lives as core-collapse supernovae. In contrast, Fe is produced in explosive burning in both supernova types \citep{ww1995}. Because of their intrinsic strength, the spectral lines of iron and magnesium are easy to measure in the spectra of late-type stars, even at low metallicity. These two elements have comparatively high cosmic abundance, and, owing to their complex atomic structure, they show many strong lines across the full stellar spectrum, from UV to the IR. Thus the observed ratio of Mg to Fe in low-mass stars, which occur in all stellar populations, is traditionally taken to be a tracer of the star formation history in galaxies \citep[][and references therein]{matteucci2014}.
 
In \citet[][, hereafter Paper 1]{bergemann2012b}, we began to systematically explore the effects of departures from 1D LTE on stellar parameters. We focussed on the non-local thermodynamic equilibrium (NLTE) spectral line formation of Fe lines using the model atmospheres derived from the mean stratification of three-dimensional (3D) stellar surface convection simulations. The effect on the determination of $\teff$, $\log g$, and metallicity [Fe$/$H]\footnote{Metallicity refers to the ratio of the stellar iron to hydrogen abundance relative to the Sun.} was quantified. In the following papers, \citet[][]{ruchti2013} and \citet[][]{hansen2013}, we applied the new NLTE methods to larger samples of stars. The last paper in the series \citep{bergemann2017}, presented the analysis of mean 3D and NLTE effects on the Mg lines in cool star spectra.

In this work, we use the methods developed in Papers 1 and 2 to carry out a detailed abundance analysis of a large stellar sample using a grid of 1D hydrostatic and mean (temporally and spatially averaged) 3D hydrodynamical models (in the following, we will refer to these stratifications as $\md$ model atmospheres) and NLTE radiation transfer. It has now been firmly established that Mg lines in spectra of late-type stars are affected by NLTE \citep{mashonkina2013}. Also, recent works have shown that the effects due to the differences between mean-3D and 1D atmospheric structures are not negligible \citep{osorio2015}. However, it is still unknown whether the combined NLTE and mean-3D effects are large enough to impact the observed $\mgfe$ distributions in Galactic populations and, consequently,  conclusions drawn from the analysis of observed datasets.

The paper is structured as follows. Section \ref{sec:observations} presents the observed stellar sample. In Section \ref{sec:methods}, we describe the details of abundance calculations. We discuss the $\feh - \mgfe$ results and compare different models in Section \ref{sec:results}. The final section \ref{sec:chemev} is devoted to the analysis of the results in the context of the chemical evolution of the Milky Way disc and halo.
\section{Observations and stellar parameters}\label{sec:observations}
\begin{figure}
\begin{center}
\includegraphics[width=0.5\textwidth, angle=0]{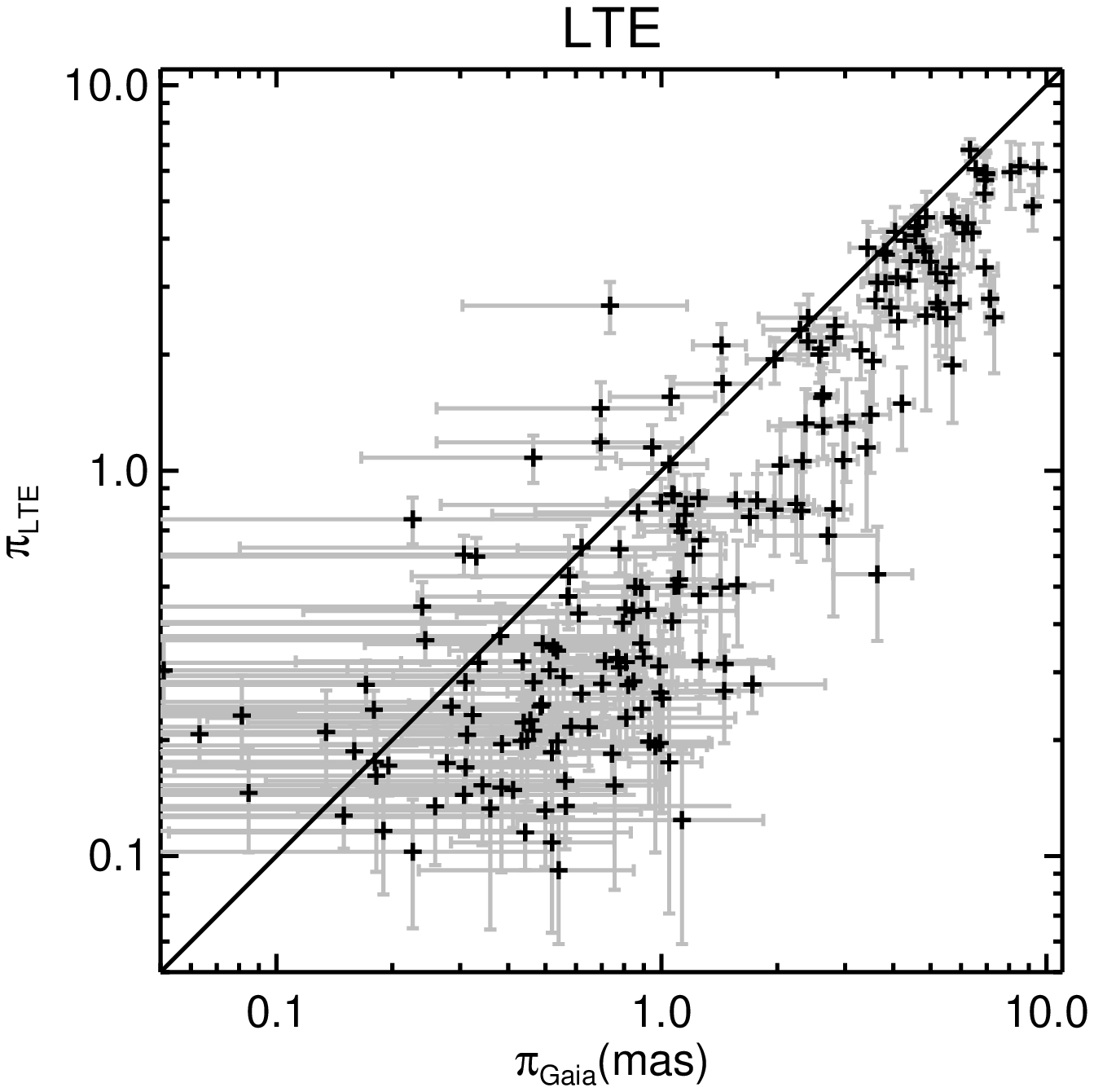}
\includegraphics[width=0.5\textwidth, angle=0]{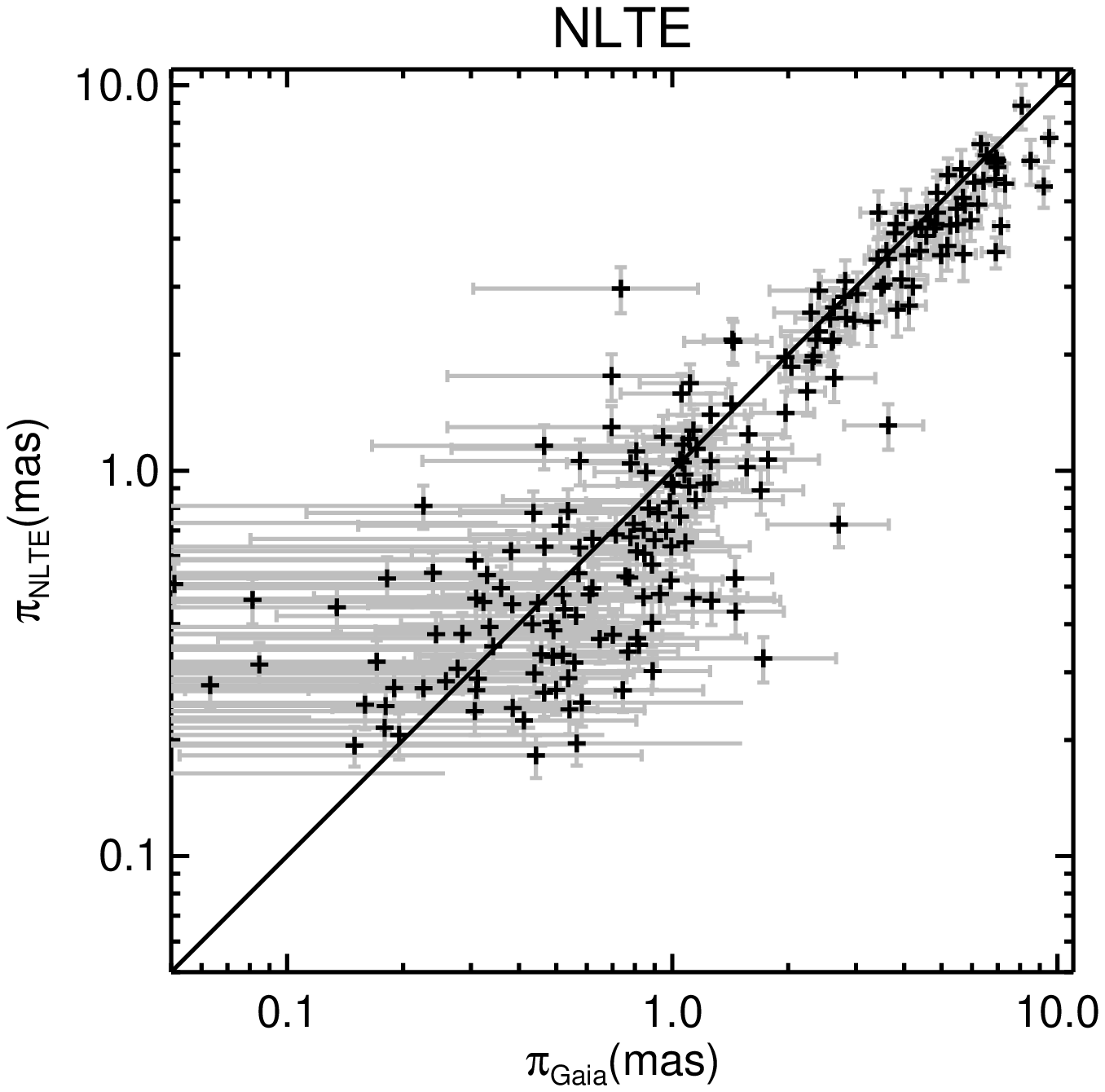}%
\caption{Comparison of spectroscopic parallax ($\pi_{\rm LTE}$, $\pi_{\rm NLTE}$) estimates for our stellar sample from \citet{ruchti2013} with the Gaia TGAS solutions  \citep{gaia1,gaia2}. Top panel: LTE stellar parameters; bottom panel: NLTE stellar parameters. Solid line represents the one-to-one relationship.} 
\label{para}
\end{center}
\end{figure}
\begin{figure}
\begin{center}
\includegraphics[width=0.5\textwidth, angle=0]{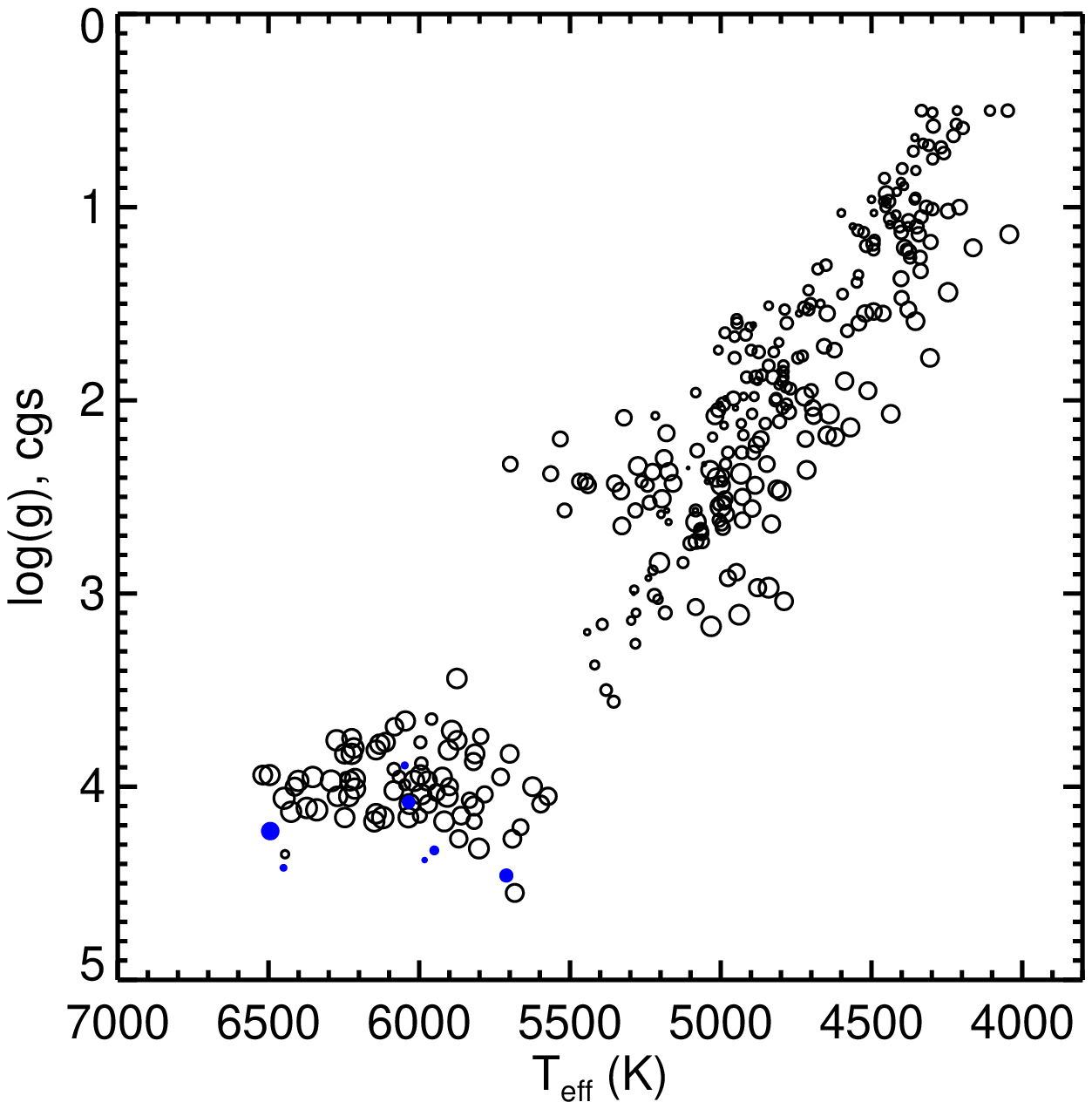}
\caption{The location of the observed sample in the $\teff$ - $\log g$ plane. NLTE stellar parameters from \cite{ruchti2013} are shown with open circles and the stars from \citet{hansen2013} with filled symbols. Symbol size is proportional to the metallicity $\feh$, which spans the range from $-3$ to $-0.5$ dex.} 
\label{hrd}
\end{center}
\end{figure}

The observed stars were taken from \citet{ruchti2011} and \citet{hansen2013}. The former sample consists of $319$ main-sequence dwarf, subgiant, and red giant stars, with metallicities ranging between ${\rm [Fe/H]}\sim-3$ up to $\sim-0.4$~dex (Fig. \ref{hrd}). The stars were selected from the RAVE Survey \citep{steinmetz2006} to study a large sample of metal-poor stars with thick disc-like kinematics and observed at high-resolution ($R>30,000$) at several facilities, including FEROS on the MPG 2.2 m telescope, MIKE spectrograph on the Magellan-Clay telescope, ARC spectrograph on the Apache Point 3.5 m telescope, and UCLES spectrograph on the Anglo-Australian telescope. All spectra, but those taken with UCLES, cover the full optical range from $\sim 3330$ \AA\ to $\sim 9380$ \AA. The UCLES data cover only the range from $4460$ to $7250$ \AA. The selection for this study was based on metallicities, [M$/$H]$_{\rm cal}  < -0.7$ \citep{zwitter2008}, distances, and 3D space motions derived using the stellar parameters in the first and second data releases from RAVE\footnote{Systematics in the RAVE stellar parameters \citep[see][]{ruchti2011,serenelli2013} resulted in the final sample containing a mixture of both disc and halo stars.}. All spectra yielded a signal-to-noise of ${\rm S/N}>80$ around $\sim6000$~\AA.
\begin{table*}
\caption{Stellar parameters and derived [Mg$/$Fe] abundance ratios for the stellar sample. Columns 2-5 give stellar parameters determined using 1D LTE excitation-ionization balance of iron. Columns 6-9 are the NLTE-opt stellar parameters \citep{ruchti2013}. Model (a) - LTE stellar parameters and LTE Mg; model (b) - NLTE-opt stellar parameters and LTE Mg; model (c) - NLTE-opt stellar parameters and NLTE Mg; model (d) - NLTE-opt stellar parameters and $\md$ NLTE Mg. The flag in column 18 gives the number of measured Mg I lines. Galactic population assignments (column 19) were taken from \citet{ruchti2011}.}
\label{table1}
\tabcolsep0.5mm
\begin{center}
\begin{tabular}{l cccc | cccc | cc cc cc cc cc}
\hline
\hline
&  \multicolumn{4}{c} {LTE-Fe}  &  \multicolumn{4}{c} {NLTE-opt}  &  \multicolumn{2}{c} {model (a)} &  \multicolumn{2}{c} {model (b)} &  \multicolumn{2}{c} {model (c)} &  \multicolumn{2}{c} {model (d)}  & Flag & Pop \\
Star  & $T_{\rm eff}$ & $\log g$ & [Fe$/$H] & $\Vmic$ & $T_{\rm eff}$ & $\log g$ & [Fe$/$H] & $\Vmic$ &  $\mgfe$  &  $\sigma$ &  $\mgfe$  &  $\sigma$ &  $\mgfe$  &  $\sigma$ &  $\mgfe$  &  $\sigma$ &  &  \\
\hline
     \object{RAVE J002330.7-163143} & 5128 &   2.40 &  -2.63 &   1.30 & 5443 &   3.20 &  -2.29 &   0.90 &   0.35  &   0.15  &   0.25  &   0.12  &   0.29  &   0.12  &   0.37  &   0.11  &  2  & 3.0    \\
     \object{RAVE J031535.8-094744} & 4628 &   1.51 &  -1.40 &   1.50 & 4774 &   2.06 &  -1.31 &   1.60 &    0.34  &   0.17  &   0.30  &   0.15  &   0.25  &   0.14  &   0.16  &   0.14  &  2  & 3.0    \\
     \object{RAVE J040840.5-462532} & 4466 &   0.50 &  -2.25 &   2.20 & 4600 &   1.03 &  -2.10 &   2.10 &    0.46  &   0.15  &   0.38  &   0.15  &   0.32  &   0.13  &   NaN  &    NaN  &  2  & 3.0 \\
     \object{RAVE J054957.6-334008} & 5151 &   2.53 &  -1.94 &   1.30 & 5393 &   3.16 &  -1.70 &   1.10 &    0.40  &   0.14  &   0.31  &   0.15  &   0.29  &   0.12  &   0.33  &   0.11  &  2  & 2.0    \\
     \object{RAVE J114108.9-453528} & 4439 &   0.50 &  -2.42 &   2.10 & 4562 &   1.10 &  -2.28 &   1.90 &    0.37  &   0.14  &   0.30  &   0.14  &   0.23  &   0.14  &   NaN  &    NaN  &  1  & 3.0   \\ 
     \object{RAVE J162051.2-091258} & 4551 &   0.67 &  -2.86 &   1.80 & 4951 &   2.04 &  -2.42 &   1.20 &   0.50  &   0.14  &   0.25  &   0.14  &   0.29  &   0.14  &   0.13  &   0.14  &  1  & 3.0   \\
     \object{RAVE J194701.6-235336} & 4562 &   1.07 &  -1.96 &   1.80 & 4797 &   1.90 &  -1.75 &   1.80 &   0.41  &   0.16  &   0.31  &   0.15  &   0.26  &   0.13  &   0.28  &   0.14  &  2  & 2.5   \\ 
\hline
\end{tabular}
\tablecomments{Table is published in its entirety in the machine-readable format. A portion is shown here for guidance regarding its form and content.}
\end{center}
\end{table*}

\begin{table*}
\caption{Stellar parameters and derived [Mg$/$Fe] abundance ratios for the stellar sample from \citet{hansen2013}. The estimates of $\log g_{\rm LTE}$ and $\feh_{\rm LTE}$ in columns 3 and 5 were computed in LTE. Columns 4 and 6 give NLTE estimates of gravity and metallicity. }
\label{table2}
\tabcolsep0.9mm
\begin{center}
\begin{tabular}{l cc cc cc cc cc}
\hline
\hline
  & & & & & &   & \multicolumn{4}{c} {[Mg$/$Fe]}  \\
\cline{8-11}
Star       & $\teff$ & log$g_{\rm LTE}$ & log$g_{\rm NLTE}$ & $\feh_{\rm LTE}$ & $\feh_{\rm NLTE}$ & $\Vmic$  & model (a) & model (b) & model (c) & model (d) \\
\hline
\object{BD $-133442$} & 6450 & 4.20* & 4.42 & $-2.56$ & $-2.46$ & 1.5 &  0.33 & 0.24 & 0.30 & 0.46 \\
\object{G 64$-$37}    & 6494 & 3.82* & 4.23 & $-3.17$ & $-3.00$ & 1.4 &  0.63 & 0.32 & 0.41 & 0.57 \\
\object{HD 3567}      & 6035 & 4.08  & 4.08 & $-1.33$ & $-1.29$ & 1.5 &  0.26 & 0.22 & 0.20 & 0.20 \\
\object{HD 19445}     & 5982 & 4.38  & 4.38 & $-2.13$ & $-2.10$ & 1.4 &  0.46 & 0.43 & 0.45 & 0.45 \\
\object{HD 106038}    & 5950 & 4.33  & 4.33 & $-1.48$ & $-1.45$ & 1.1 &  0.49 & 0.45 & 0.43 & 0.44 \\
\object{HD 121004}    & 5711 & 4.46  & 4.46 & $-0.73$ & $-0.71$ & 0.7 &  0.25 & 0.24 & 0.23 & 0.27 \\
\object{HD 122196}    & 6048 & 3.89  & 3.89 & $-1.81$ & $-1.75$ & 1.2 &  0.28 & 0.23 & 0.25 & 0.27 \\
\hline
\end{tabular}
\end{center}
\end{table*}

We assume two different sets of input stellar parameters: 1D LTE and NLTE-opt, as described in \citet{ruchti2013}. 1D LTE values were determined by the classical method of LTE excitation-ionisation balance of Fe lines. Our second input dataset is NLTE-opt, for which accurate estimates for the effective temperature (T$_{\rm eff}$), surface gravity ($\log~g$), and metallicity of the stars were determined as follows. For $\teff >$ 4500 K, the NLTE-opt temperature was estimated from the combination of fits to the wings of Balmer lines and the photometric $\teff$ calibration from \citet{ruchti2011}. Below $4500$ K, the \citet{ruchti2011} calibration was adopted. The surface gravity, metallicity, and micro-turbulence were estimated through the NLTE ionisation balance of \fei\ and \feii\ lines. The adopted stellar parameters are listed in Table \ref{table1}.

The second sub-sample\footnote{\object{BD-133442}, \object{G 64-37}, \object{HD 3567}, \object{HD 19445},  \object{HD 106038}, \object{HD 121004}, \object{HD 122196}.} was selected from \citet{hansen2012,hansen2013} and includes $7$ dwarfs with a wide range of metallicities reaching [Fe$/$H] $\sim -3$ and $-0.7$ that belongs to the halo and disc, respectively. Most of these stars have been observed with UVES/VLT in settings U-564, U-580, U-800, and U-860. This ensures a wavelength coverage that contains all the Mg lines under study while three of the stars have been observed using HIRES/Keck covering 3120 - 4650\,\AA. The slit width used is between 0.7 to 1.0 arcsec which ensures that all spectra are of high resolution ($R\sim 40\,000 - 62\,000$). Most spectra are of high signal-to-noise ratio (SNR), with typical values of $\sim200$ at 5100 \AA. For most of the stars the stellar parameters have been determined using photometry (IRFM) and parallax \citep[see][for details]{hansen2012}. In a few cases (marked by 'c' in Table B.1 in \citealt{hansen2012}) we relied on stellar spectra to derive temperature and gravity owing to larger uncertainties in parallax and colour. These stars had their parameters corrected and recomputed using NLTE corrections for Fe. Table \ref{table2} gives the adopted stellar parameters for the \citet{hansen2013} sample.

We performed several tests in \citet{ruchti2013}, showing that the $\teff$ measurements were in agreement with the photometric, interferometric, and the infra-red flux method temperatures. The accuracy of NLTE-opt surface gravities was tested by comparing them with astrometric (Hipparcos) and asteroseismic gravity estimates. For $214$ stars in our sample, parallax estimates have become available from the Gaia Data Release 1 (Tycho-Gaia Astrometric Solution, TGAS) \citep{gaia1, gaia2}. Figure \ref{para} compares our spectroscopic parallax estimates to the astrometric solution from TGAS, using either the 1D LTE or the NLTE-opt estimates. The NLTE-opt spectroscopic parallaxes are in good agreement with the TGAS data; stars with small parallaxes (luminous red giants beyond $\sim 2$ kpc) deviate from the line due to their large uncertainties in the astrometric parallax. 

\begin{figure}
\begin{center}
\includegraphics[width=0.5\textwidth, angle=0]{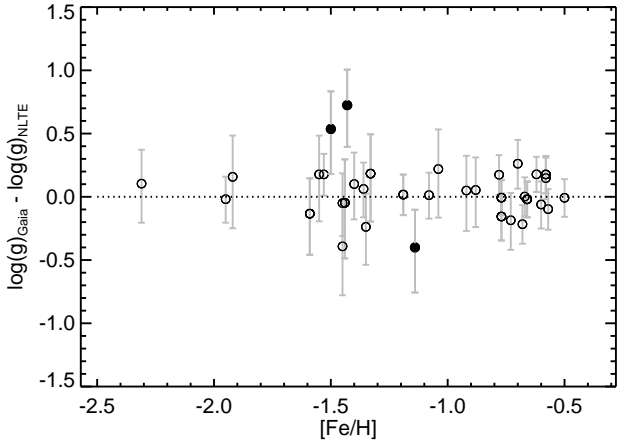}
\caption{Difference between the astrometric estimate of surface gravity and the spectroscopic NLTE estimate for the stars with parallax errors within $30 \%$ and extinction less than $0.1$. Candidate binaries or multiple systems are shown with filled circles.  The error bars represent the combined uncertainties of the astrometric and spectroscopic $\log g$ estimates; typically, they are dominated by the errors of the TGAS parallaxes.} 
\label{logg}
\end{center}
\end{figure}

Gaia parallaxes can be used to test the accuracy of the adopted surface gravity scale. Errors in TGAS dataset are very large for most stars in the sample therefore we limit the comparison to those objects for which parallaxes are known to better than $30\%$. Figure \ref{logg} shows the comparison of the $\log g$ estimates based on NLTE-opt method and the astrometric estimates derived using the TGAS parallaxes. The astrometric gravities are computed using the standard relationships:
\begin{equation}
M_{\rm V} = m_{\rm V}  + 5 \log (\pi) + 5 - A_{\rm V}
\end{equation}
\begin{equation}
M_{\rm bol} = M_{\rm V} + {\rm BC}
\end{equation}
\begin{equation}
\log L  = -0.4 (M_{\rm bol} - M_{\rm bol, \sun})
\end{equation}
\begin{equation}
[g] =  [M]  - [L] + 4[\teff]
\end{equation}
where the square brackets represent the logarithmic ratio of a parameter with respect to the Sun, $m_{\rm V}$ is the apparent visual magnitude, $M$ is the mass of a star taken from \citet{serenelli2013}, $BC$ the bolometric correction, and $L$ the luminosity. We adopt the photometry from \citet{munari2014}, and extinction in the V-band, $A_{\rm V}$, is computed using the coefficients in \citet{schlafly2011}. Bolometric corrections are taken from \citet{casagrande2014}. The plot shows that there is a good agreement between the astrometric and spectroscopic NLTE-opt gravities. The few outliers appear to be binaries or stars in multiple systems, according to the large astrometric excess noise in the Gaia DR1.

Metallicities, even though based on 1D NLTE analysis, are expected to be sufficiently accurate for our purposes, and we do not recompute them. In \citet{bergemann2012b}, we showed that the difference between the 1D NLTE and $\md$ NLTE metallicities is very small, and is of the order $0.01 - 0.04$ dex for the full range of stellar parameters analysed here. In this work, metallicities are used to compute the model atmospheres for Mg abundance determinations and then to calculate $\mgfe$ abundance ratios.
%
%
%
\section{Methods}{\label{sec:methods}}
\subsection{Model atmospheres and linelist}{\label{sec:inputdata}}
We use two different sets of model atmospheres. The reference 1D models are hydrostatic plane-parallel MAFAGS-OS model atmospheres \citep{grupp2004a,grupp2004b}. Mean 3D models were derived by averaging the physical structure from 3D {\sc Stagger} stellar surface convection simulations \citep{collet2011,magic2013}. A description of the averaging procedure of time-dependent 3D model atmospheres is given in \citet{bergemann2012b}. The applicability of mean 3D models in spectroscopic analysis was studied in Paper 1 using the full 3D NLTE radiative transfer for selected Mg spectral lines.

With the goal to establish reliable abundances, we have taken special care to select only the most robust \mgi\ spectral lines. These are the lines that are not excessively affected by blending, are not located in regions with poorly-defined quasi-continuum, and are visible at low metallicity. Also in Paper 1, we showed that the 5711 \AA\ and, to a lesser degree, the 5528 \AA\ features should be preferred for abundance analyses in cool type stars, because other features (including the resonance line at 4571 \AA\ and the optical triplet at 5172 and 5183 \AA) are more sensitive to the structure of stellar atmospheres, and thus to the limitations caused by the use of the hydrostatic models. Therefore in this paper, we limit the analysis to the spectral lines at 5711 \AA\ and 5528 \AA. The atomic data are those given in Paper 1: $\log gf$ (5711 \AA) $= -1.742$, and $\log gf$ (5528 \AA) $= -0.547$; both estimates of transition probabilities are adopted from \citet{pehl2017}. The damping caused by inelastic collisions with hydrogen atoms is from \citet{barklem2000} for the 5528 \AA\ line; for the 5711 \AA\ line, we use the damping constants kindly provided by P. Barklem (priv. comm.).

The non-LTE statistical equilibrium is computed using the DETAIL code \citep{butler1985} and the updated Mg atomic model by \citet{bergemann2015} which builds upon the model by \citet{mashonkina2013}. Line profile fitting and abundance determinations are performed using the spectrum synthesis code SME \citep{valenti2012}. When fitting each \mgi\ line, we allow for individual variations of macroturbulence. However, the $\Vmac$ estimates for both \mgi\ lines are not very different and only show a small systematic offset of $\sim 1$ kms$^{-1}$ that is likely caused by the differences in the velocity field at the depth of line formation. We assure the quality of all line fits by visual inspection. We also determine the line equivalent widths (EW) through the integration of the observed line profiles and of the best-fit models. This procedure is very useful to eliminate poor fits, caused, for example, by line veiling or contamination by blends. Furthermore, all measurements, for which the EW is less than 5 m\AA\ or the difference between the observed and synthetic EW is larger than $3 \%$ are disregarded. The EW measurements are given in Table \ref{table3}.
\begin{table}
\caption{Measurements of the equivalent widths of the two \mgi\ diagnostic lines for the program stars. The equivalent widths are given in m$\AA$.}
\label{table3}
\tabcolsep1.1mm
\begin{center}
\begin{tabular}{l rr}
\hline
\hline
        & \multicolumn{2}{c} {EW}  \\
\cline{2-3}
Star  &  \mgi~5528 & \mgi~5711 \\
\hline
 \object{RAVE J002330.7-163143} & 54.96 &   5.45 \\
 \object{RAVE J031535.8-094744} &159.80 &  71.13 \\
 \object{RAVE J040840.5-462532} &117.82 &  29.01 \\
 \object{RAVE J054957.6-334008} & 99.60 &  23.14 \\
 \object{RAVE J114108.9-453528} & 90.74 &  14.48 \\
 \object{RAVE J162051.2-091258} & 60.12 &   2.34 \\
 \object{RAVE J194701.6-235336} &128.36 &  41.42 \\
\hline
\end{tabular}
\tablecomments{Table is published in its entirety in the machine-readable format. A portion is shown here for guidance regarding its form and content.}
\end{center}
\end{table}

\subsection{Abundance determination}\label{sec:nltecorr}
The Mg abundances are computed using two different sets of stellar parameters, 1D LTE and NLTE-opt (Sect. \ref{sec:observations})

\begin{figure*}
\begin{center}
\hbox{
\includegraphics[width=0.33\textwidth, angle=0]{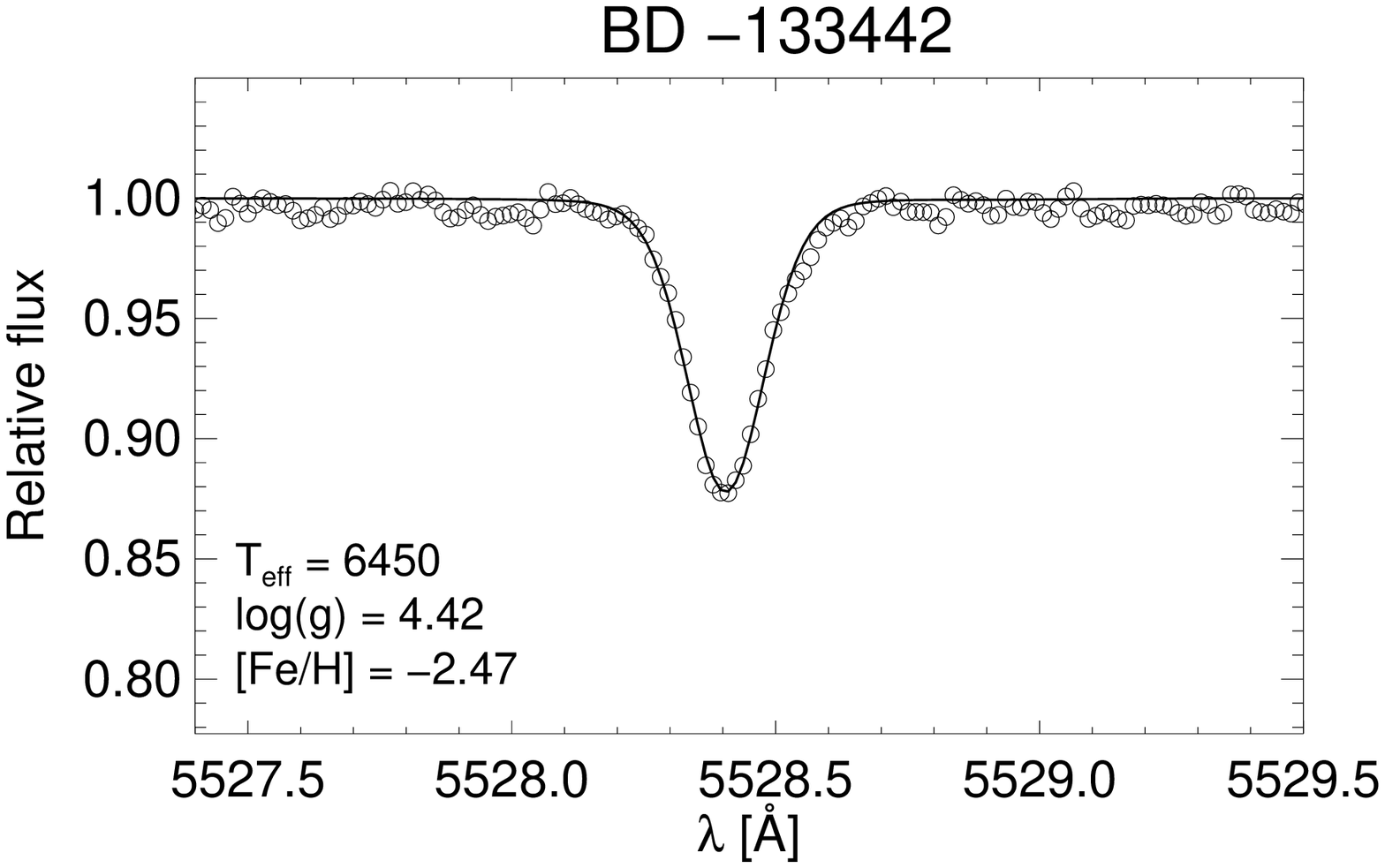}
\includegraphics[width=0.33\textwidth, angle=0]{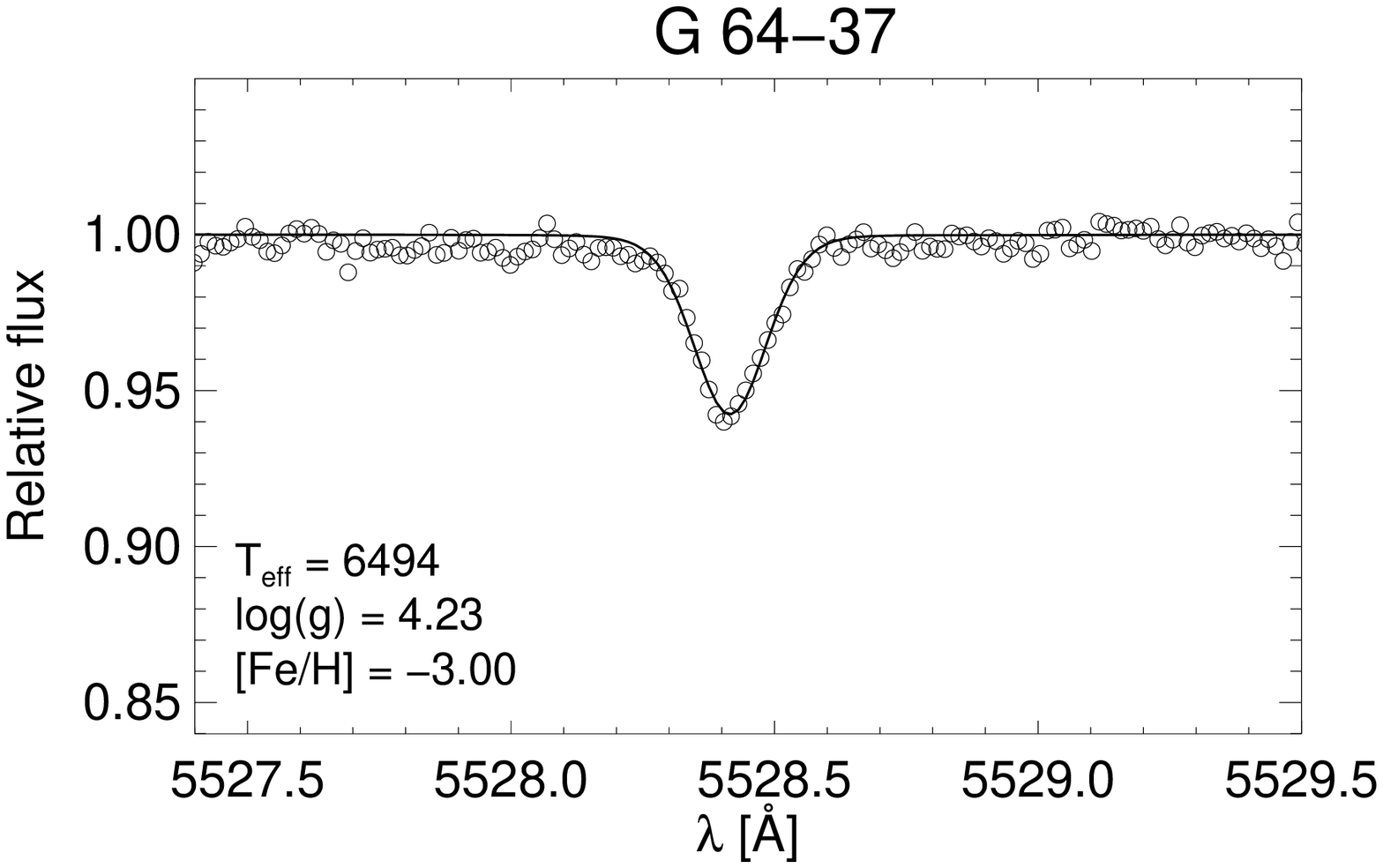}
\includegraphics[width=0.33\textwidth, angle=0]{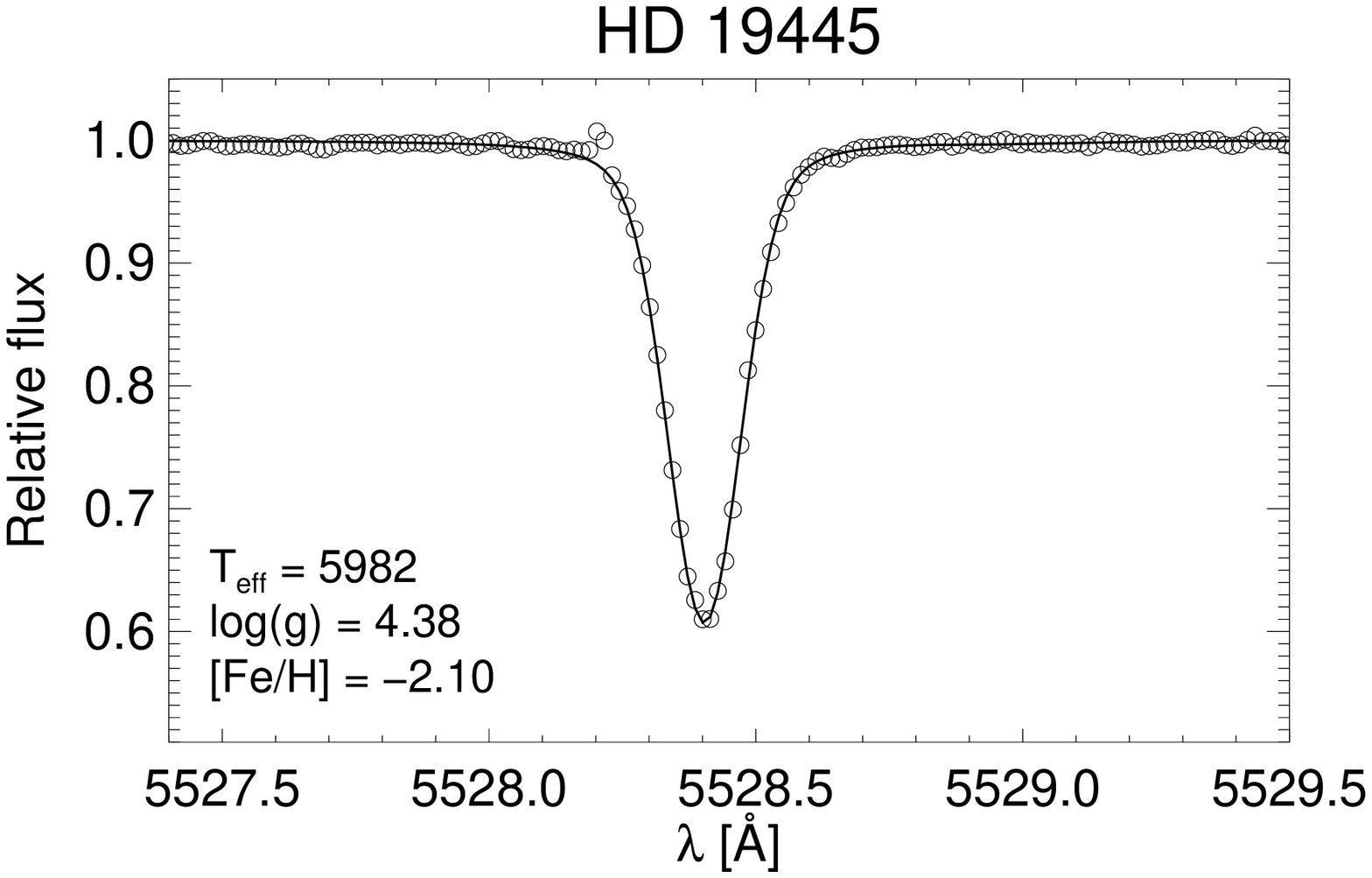}
}
\hbox{
\includegraphics[width=0.33\textwidth, angle=0]{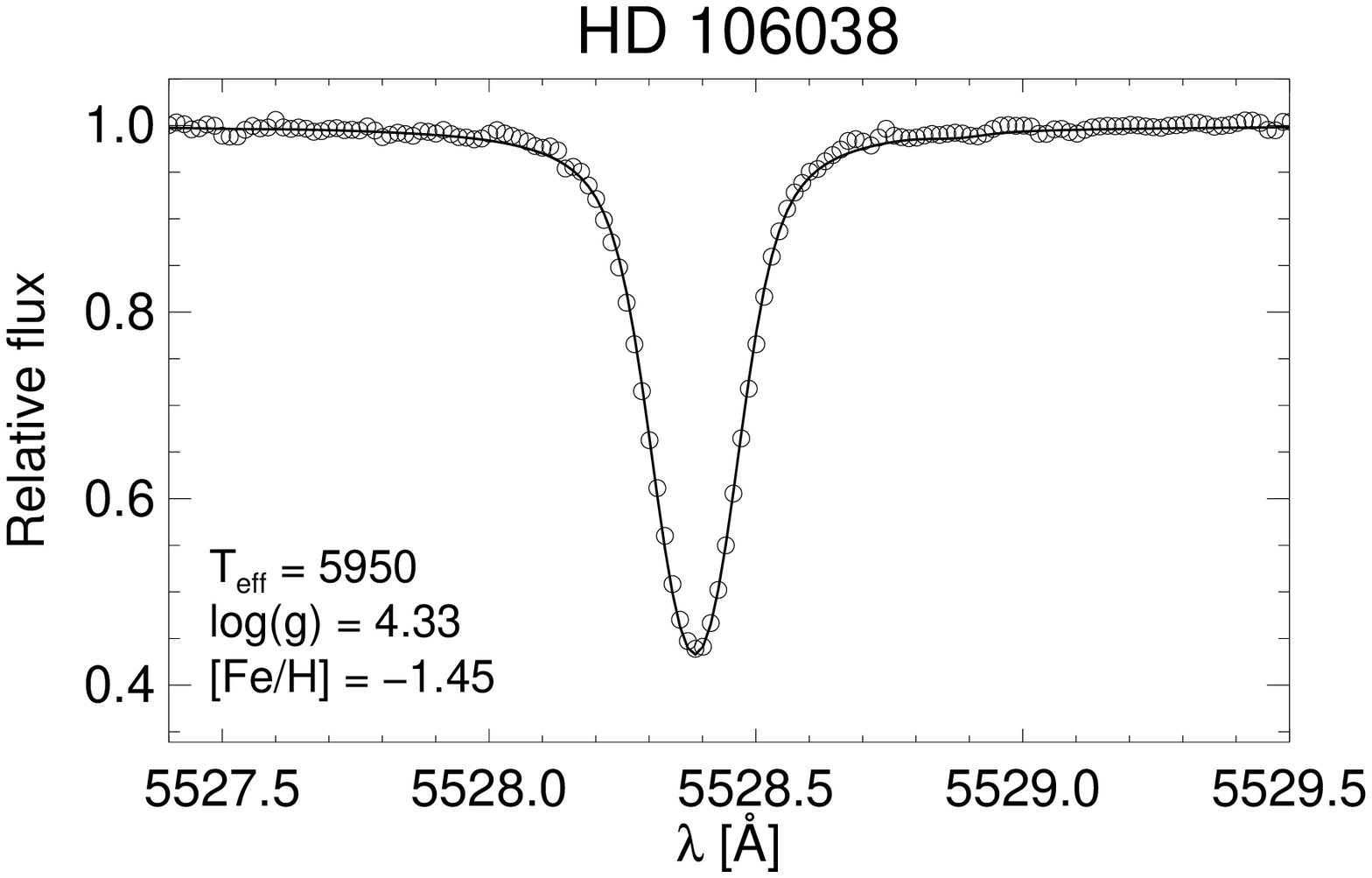}
\includegraphics[width=0.33\textwidth, angle=0]{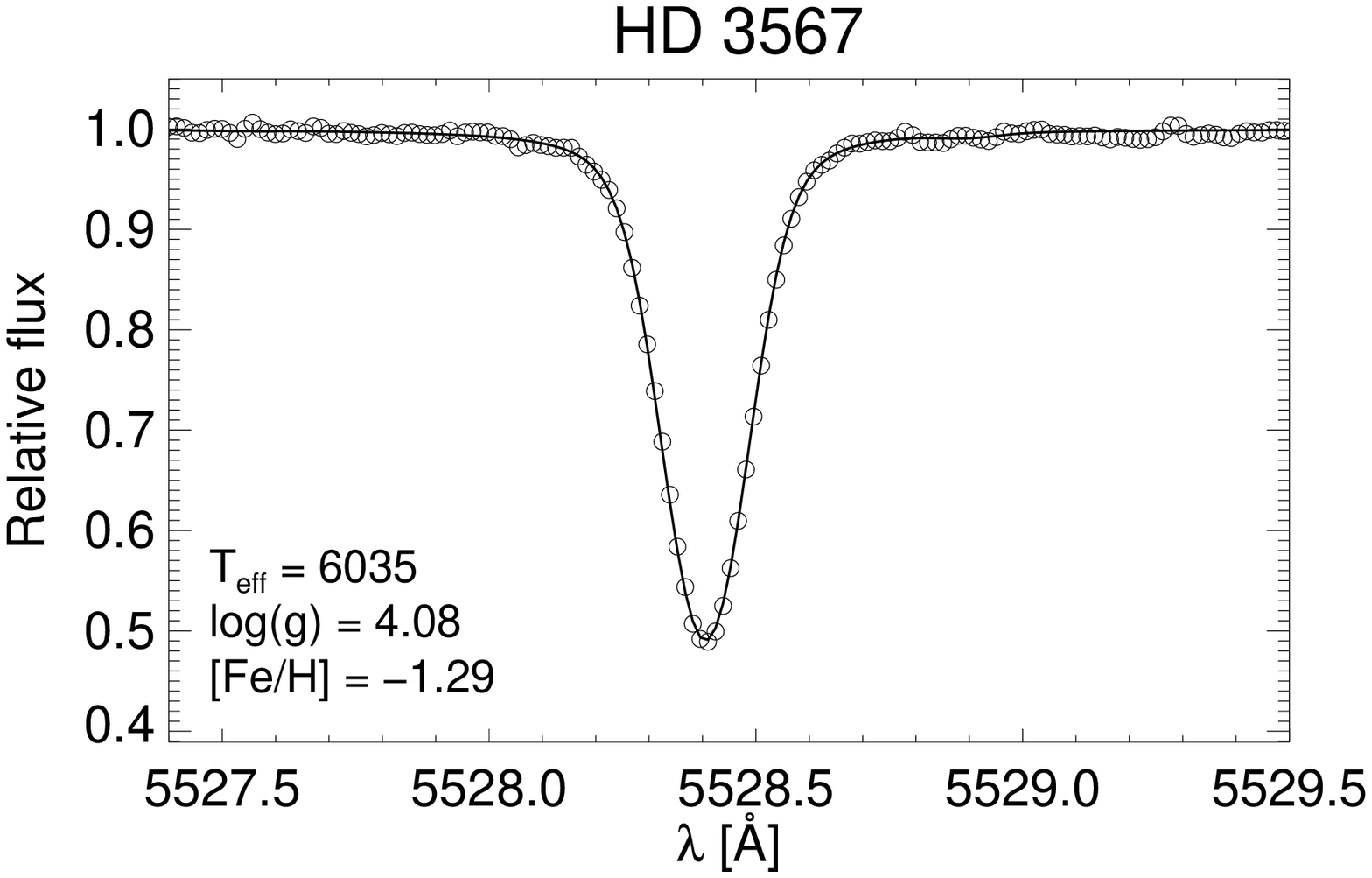}
\includegraphics[width=0.33\textwidth, angle=0]{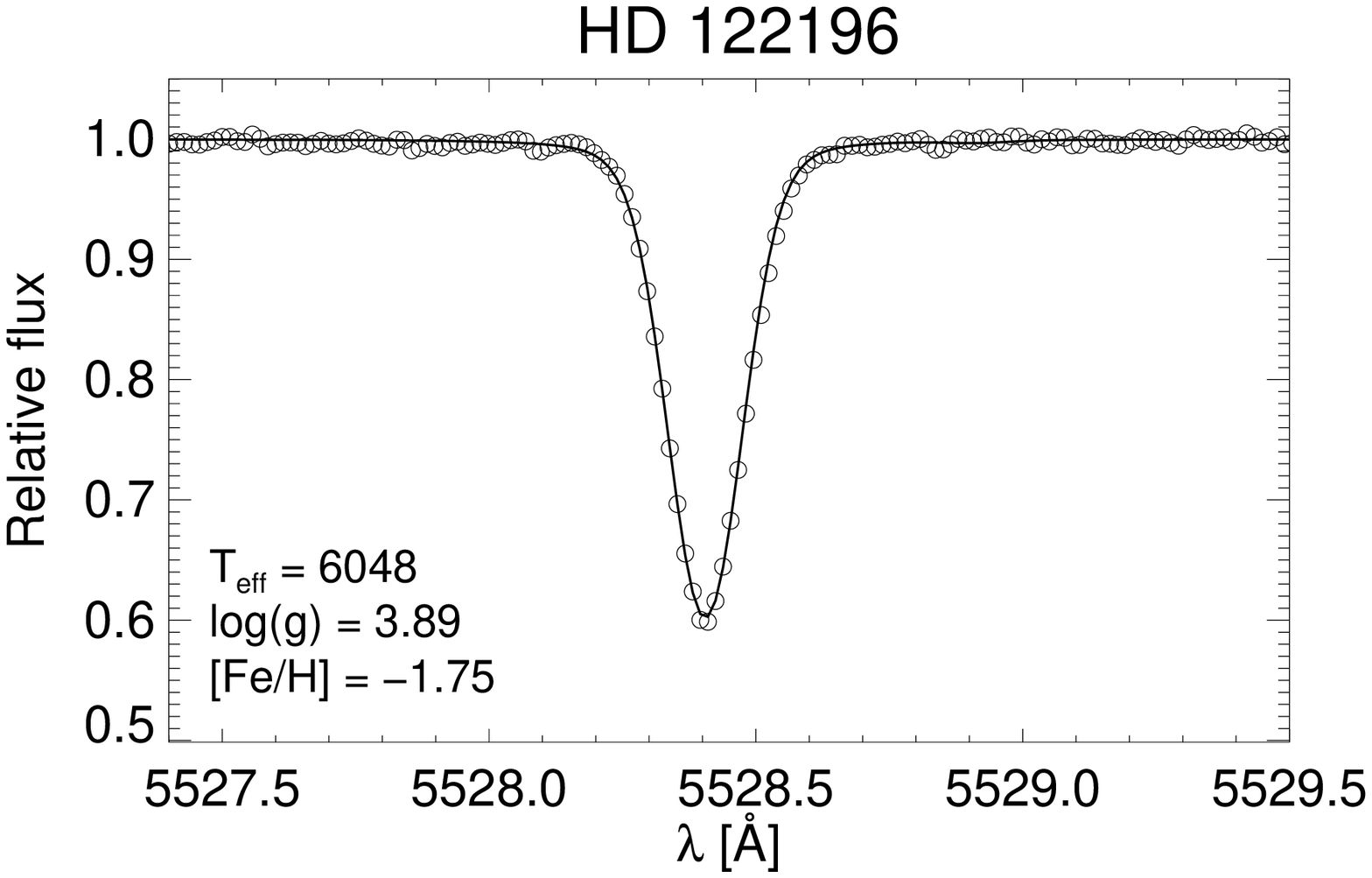}
}
\caption{The observed profiles (open circles) and best-fit LTE profiles (solid line) for the 5528 \AA\ \mgi\ line for the stars from the \citet{hansen2013} sample.}
\label{profiles}
\end{center}
\end{figure*}

To derive 1D and $\md$ NLTE Mg abundances, we first compute LTE Mg abundance for each \mgi\ line using our best NLTE-opt stellar parameters. Some examples of the best-fit line profile fits are shown in Fig. \ref{profiles}.
Then we compute the 1D NLTE and $\md$ NLTE Mg abundance corrections separately, and apply the corrections to the LTE Mg abundances. The abundance correction is defined as the difference between the abundance of a spectral line obtained using the NLTE calculations (with either 1D hydrostatic or $\md$ model atmospheres) and the one obtained in 1D LTE (both sets computed with the more accurate NLTE-opt stellar parameters), and are hereafter denoted by:
\begin{equation}
\Delta \rm A (\rm Mg) =  \rm A (\rm Mg )_{\rm{NLTE}} - \rm A (\rm Mg)_{\rm{LTE}}
\end{equation}
Since the corrections are also sensitive to the atmospheric abundance of the element, we compute the correction relative to the Mg abundance found in our LTE calculations, not relative to the scaled-solar abundance of the element.

In 1D, the NLTE corrections typically range from $-0.1$ to $0.1$ dex. Only for the model atmospheres with the lowest surface gravity ($\log g < 1$ dex) and effective temperature ($\teff < 4500$ K), the 5711 and 5528 \AA\ lines show a correction of $-0.15$ dex. The $\md$ NLTE corrections are larger, ranging from $-0.2$ (cool low-gravity models) to $0.2$ (metal-poor models of turnoff stars with $\teff > 6500$ K) dex. Figures \ref{abcor1} and \ref{abcor2} (in the Appendix) illustrate this behaviour for the both types of models as a function of $\teff$, $\logg$, and $\feh$.

The final Mg abundance is computed as the average of the measurements based on the 5711 and 5528 \AA\ features. The uncertainties of the Mg measurements are derived as follows. The systematic errors are computed by propagating the uncertainties in $\teff$, $\log g$, [Fe$/$H], and atomic data (transition probabilities) in the Mg abundance error, also including the $0.2$ kms$^{-1}$ error in micro-turbulence. The systematic error was computed by summing the error components in quadrature:
\begin{equation}
\sigma_{\rm tot} = (\sigma_{\rm \Vmic}^2 + \sigma_{\rm \feh}^2+ \sigma_{\log gf}^2 + \sigma_{\rm \log g}^2 + \sigma_{\teff}^2)^{1/2}
\end{equation}

In this approach, the uncertainties in stellar parameters are assumed to be uncorrelated, which may not be true as shown by \citet{schoenrich2014}. However, the full probability distribution in parameter space is not available to us. A full treatment is beyond the scope of this paper, and in this case not necessary for our general conclusions.
The internal error is given by the standard deviation of the measurement based on the 5711 and 5528 \AA\ lines. If only one \mgi\ line is available (Table \ref{table3}), the internal uncertainty is taken to be $0.05$ dex. Then the total error is computed as the sum of the statistical and systematic error components.
%
%
%
\section{Results}{\label{sec:results}}
\begin{figure*}
\begin{center}
\gridline{\fig{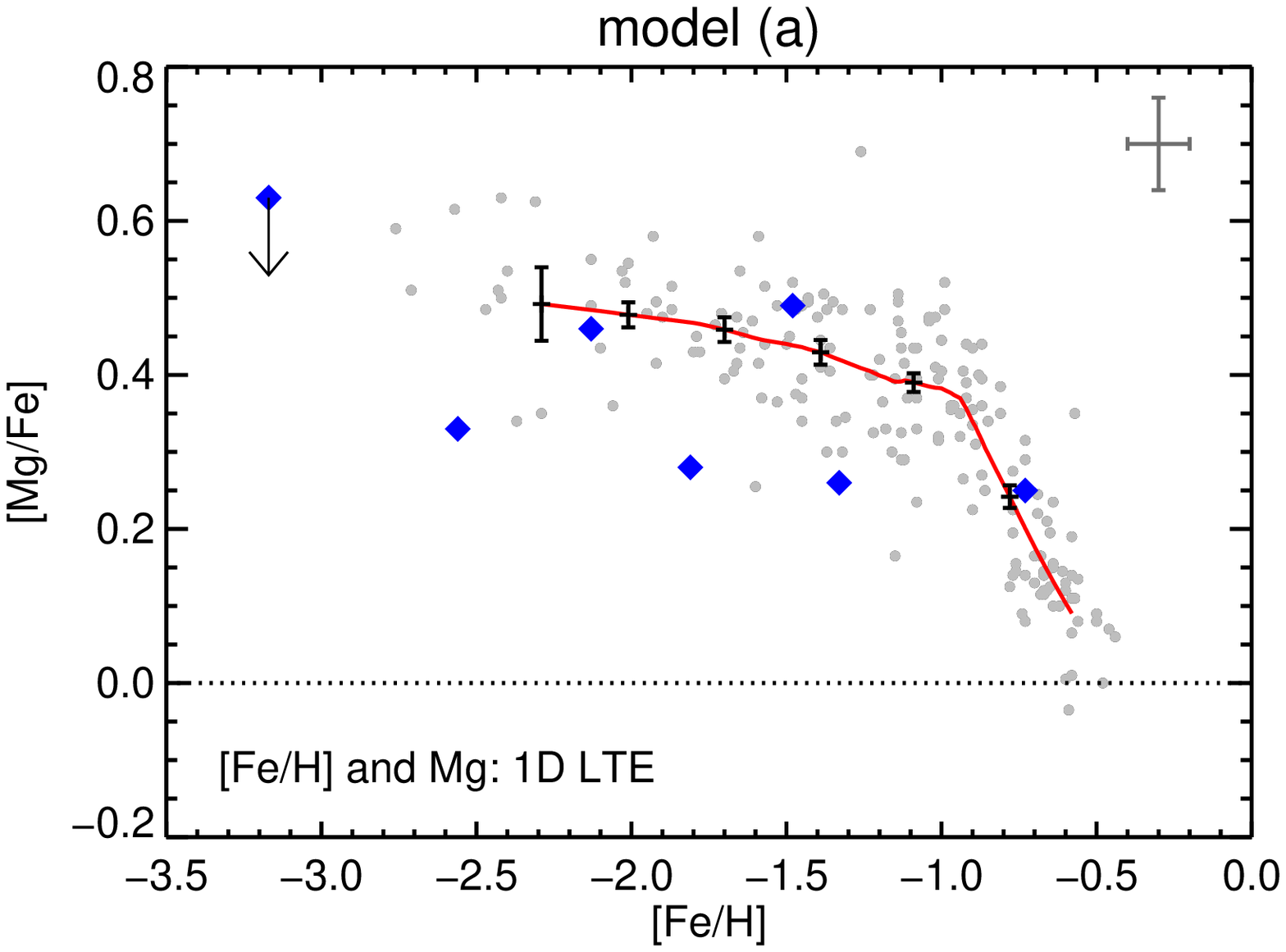}{0.5\textwidth}{(a)}
              \fig{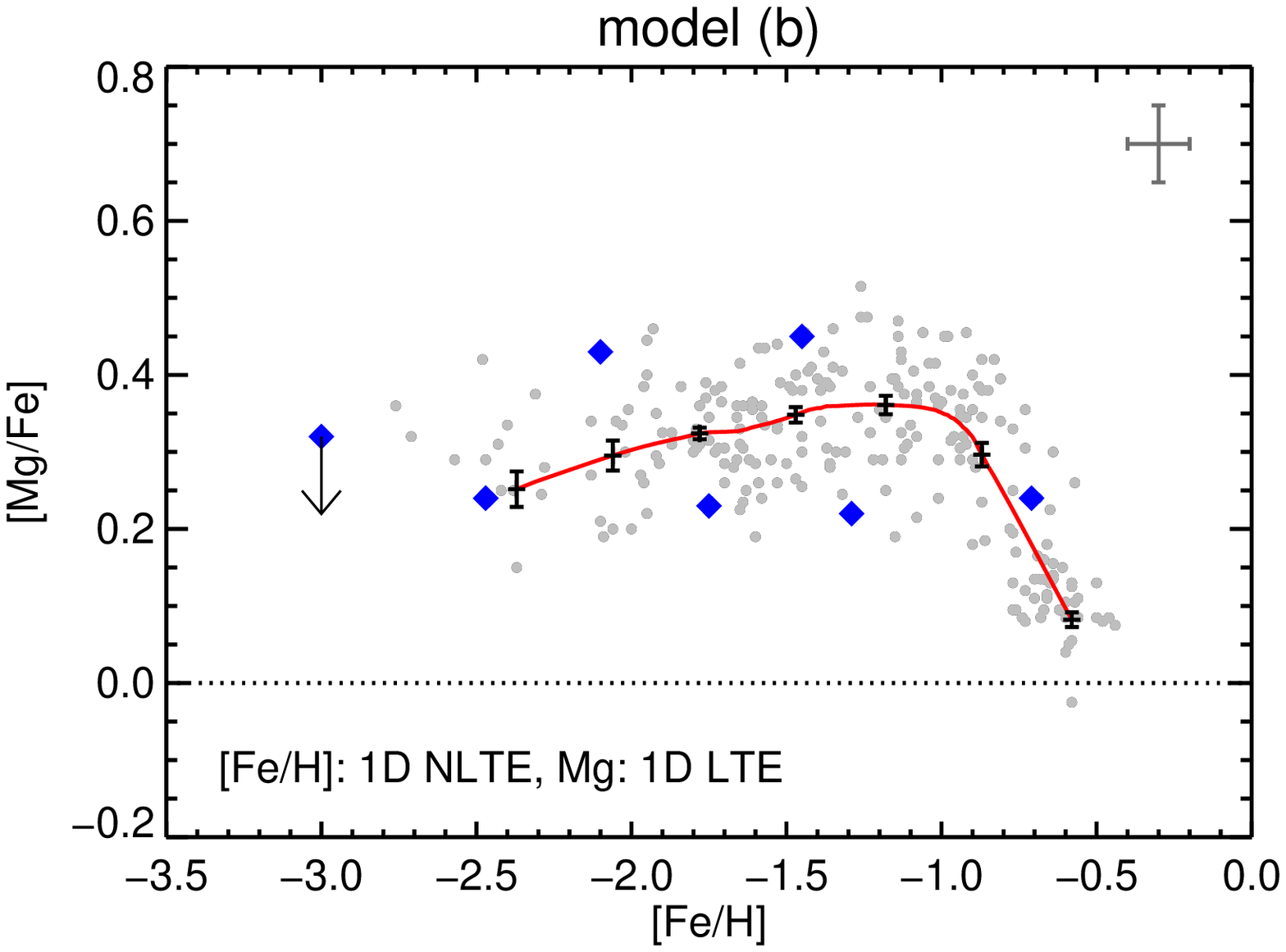}{0.5\textwidth}{(b)}}
\gridline{\fig{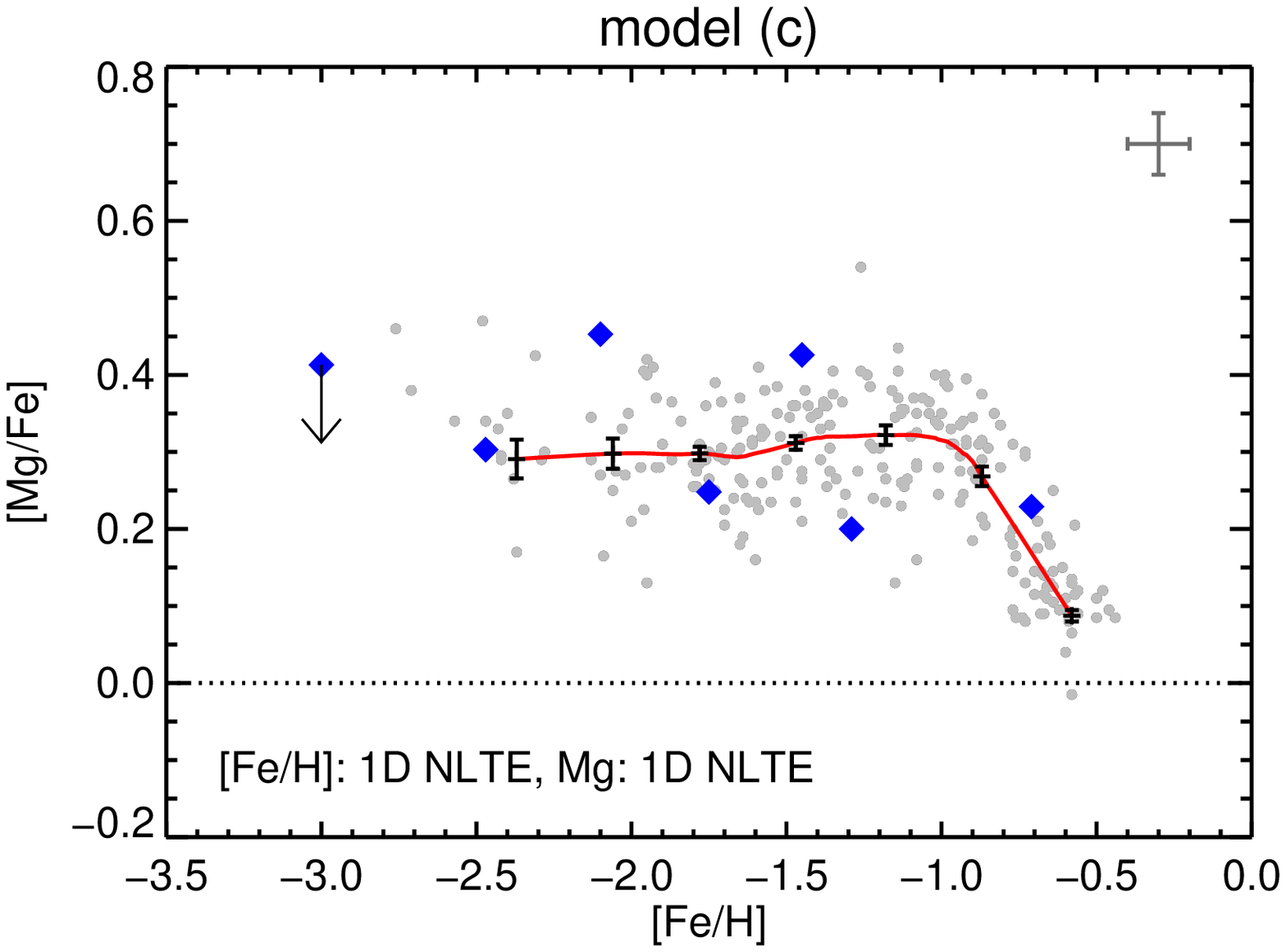}{0.5\textwidth}{(c)}
              \fig{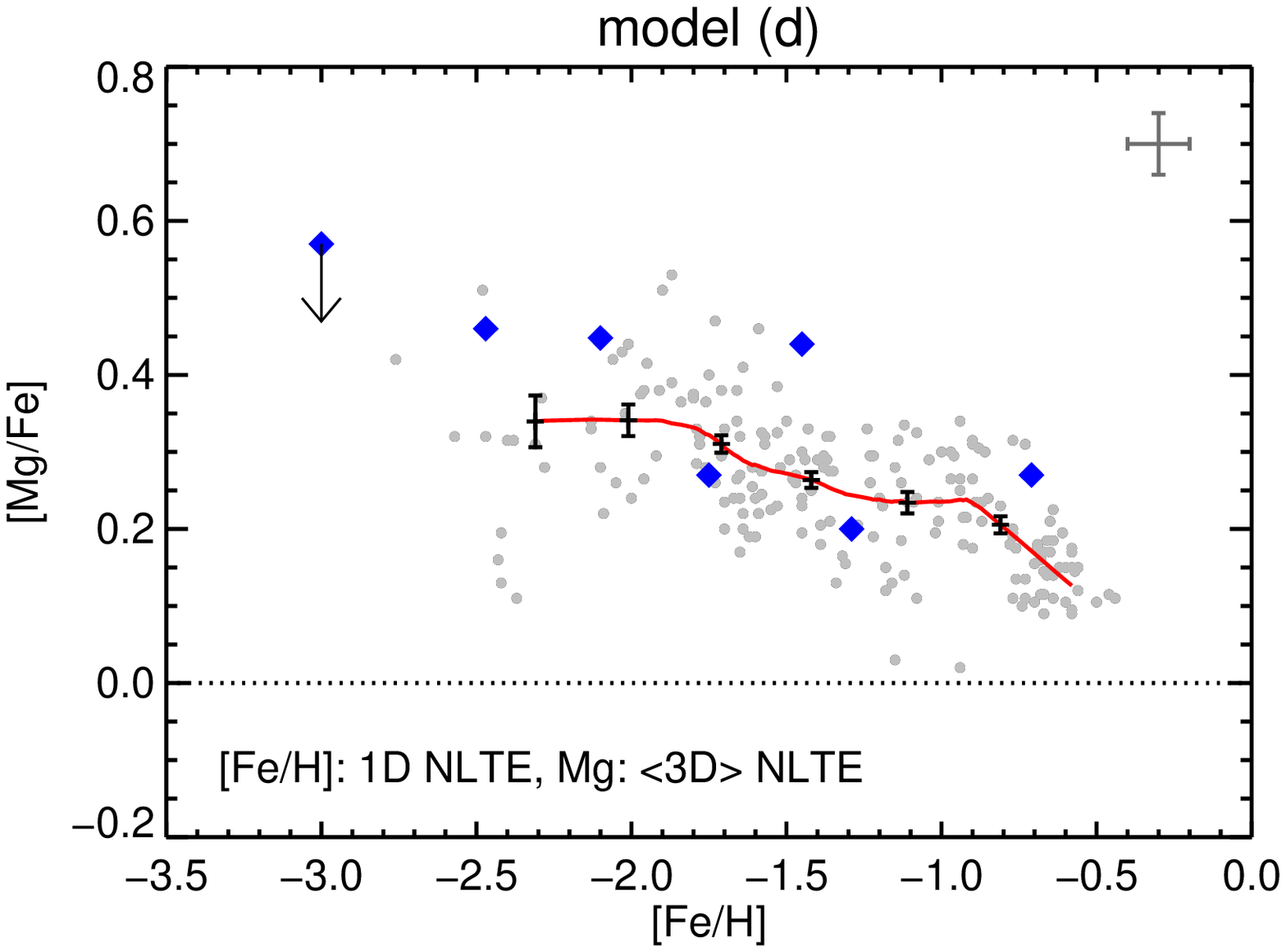}{0.5\textwidth}{(d)}}
\caption{[Mg$/$Fe] abundance ratios as a function of metallicity determined using the four modelling scenarios: model (a) - LTE stellar parameters and LTE Mg; model (b) - NLTE-opt stellar parameters and LTE Mg; model (c) - NLTE-opt stellar parameters and NLTE Mg; model (d) - NLTE-opt stellar parameters and $\md$ NLTE Mg. The number of measurements shown in the panel (d) differs from the other panels, because we could not achieve convergence  of NLTE level populations in the $\md$ grid for some very cool and low-gravity models. The reference main-sequence stars from \citet{hansen2013} are shown with filled blue diamonds. The error bar in the top right corner is the internal uncertainty of the measurements. The red line depicts the smoothed $\mgfe$ distribution computed using LOESS regression. The errors bars attached to the LOESS curve represent the sample estimated uncertainty of the mean, i.e. the sample dispersion divided by $\sqrt{N-1}$, with N the number of stars in the metallicity bin of $0.3$ dex.}
\label{final}
\end{center}
\end{figure*}

The derived Mg abundances are given in Tables \ref{table1} and \ref{table2} and are shown in Figure \ref{final} as a function of metallicity. The error bar in the top right corner reflects the internal uncertainty (line-to-line dispersion) of the measurements. For giants with $\teff < 4500$ K or surface gravity $ < 1.5$, we cannot provide a robust estimate of the systematic uncertainty, because the original $\md$ model atmosphere grid does not sample this regime of stellar parameter space.  Also for some very cool and low-gravity models we could not achieve convergence of NLTE level populations in the $\md$ grid. These stars are, therefore, not shown in Figure \ref{final}. 

The LTE $\mgfe$ - [Fe$/$H] distributions computed using 1D LTE stellar parameters are shown in Fig. \ref{final}a. The other three panels (b,c,d) show Mg abundances computed using the more accurate NLTE-opt stellar parameters (see Sect. \ref{sec:observations}), with 1D LTE, 1D NLTE, and $\md$ NLTE approach for Mg abundance determinations. To make the average trend visible, Figure \ref{final} also shows the smoothed [Mg$/$Fe] distributions computed using the robust form of the LOESS regression \citep{cappellari2013}. The smoothing parameter is set to $0.4$. We ignore ten data points at the low- and high-metallicity end, in order to minimise the error caused by the data sparseness close to the edges of the distribution. The errors bars on the LOESS curve represent the sample estimated uncertainty of the mean computed in $0.3$ dex metallicity intervals. 

Comparison of the models (a) and (b) in Figure \ref{final} clearly illustrates the important effect of stellar parameters ($\teff, \log g, \feh$) on the distribution of stars in the $\feh$ - $\mgfe$ plane. Assuming 1D LTE in stellar parameter and abundance determinations (Fig. \ref{final}a), we find that the shape of the $\mgfe$ distribution with metallicity has two components. In the metal-poor stars, $\mgfe$ ratio mildy increases with decreasing metallicity. At metallicites higher than $\feh \approx -0.8$ dex, the $\mgfe$ ratio drops towards the solar value. The change of the slope occurs at slightly lower metallicity, compared to what is known in the literature \citep{nissen1994,fu1998,fu2004,bensby2014,kordopatis2015}, but this is simply because of the metallicity selection in our sample that creates a strong bias against metal-rich stars with $\feh \gtrapprox -0.7$. In other samples, which are more complete in the metallicity regimes probing the thin disc \citep[e.g.][]{bensby2014}, the $\mgfe$ distribution changes slope at $\feh \sim -0.6$. 

If we adopt more accurate NLTE-opt stellar parameters, but assume LTE line formation for Mg (Fig. \ref{final}b), the location of the knee in the $\feh$-[Mg$/$Fe] does not change, but the slope of the metal-poor part of the $\mgfe$ distribution changes. For the most metal-poor stars ($-3 \leq \feh \leq -1.5$), $\mgfe$ is an increasing function of metallicity. This effect is the consequence of the NLTE effect on iron abundances \citep[see][for more details]{ruchti2013} that implies that the overall distribution becomes more metal-rich. The [Mg$/$Fe] estimates come out lower, most notably in the metal-poor regime, where the NLTE effects on iron abundance are significant.

The NLTE Mg abundances derived using the NLTE-opt stellar parameters (Fig. \ref{final}c) are not too different from the model b, with $\mgfe$ only slightly deviating from the plato at $0.3$ dex below $\feh \sim -1$. The scatter of $\mgfe$ is relatively constant with metallicity, of the order $\sim 0.15$ dex, slightly larger than the internal uncertainty of the abundance measurements, however, the number of stars deviating from the trend line increases with decreasing $\feh$. Some of these stars have very low $\mgfe$, close to the solar value, and a few very metal-poor objects have $\mgfe$ ratios higher than the mean trend by $\gtrsim 0.2$ dex. 

The $\md$ NLTE (Fig. \ref{final}d) $\mgfe$ distribution is quite different from the distribution obtained using model (c). First, the $\mgfe$ ratio does not reach the $0.4$ dex enhancement at the location of the break at $\feh \sim -1$, but remains at the level of $0.25$ dex. Beyond this point, the $\mgfe$ ratio appears to grow monotonically towards lower metallicities. There is an indication for the $\mgfe$ ratio stalling or even falling beyond $\feh \sim -1.8$, however the increasing scatter and dwindling stellar numbers make this change barely significant, as indicated by the black error bars. We will see in the next section that this trend harbours the metal-poor thick disc tail with actually constant $\mgfe$, overlaid by the halo population exhibiting a strong trend. At even lower metallicity, the star-to-star scatter increases and, again, we find a number of stars, which are more enriched / or depleted in $\mgfe$ ratios than the trend line. Whereas our $\md$ NLTE $\mgfe$ distribution contrasts with our 1D LTE/NLTE results and with other optical studies of Mg abundances in the disc stars, it is consistent with the results of the APOGEE survey \citep{holzmann2015,garcia2016} based on the infrared H-band stellar spectra. They also find maximum enrichment of $\mgfe$ $\sim 0.25$ dex at the metallicities of the thin and thick disc, although the analysis is based on 1D LTE.  
\section{Interpretation and Implications for Chemical Evolution}{\label{sec:chemev}}
The differences between the approaches to spectroscopic Mg abundance determinations have implications for Galactic chemical evolution. As an $\alpha$-capture element (proton and neutron numbers are a multiple of helium), \mgi\ is thought to originate from essentially the same production sites as the other $\alpha$-elements, i.e. mostly ejected from massive stars in their core collapse, while Fe is produced in both SN Ia and in SN II. The time delay of SNIa sets a natural clock \citep{tinsley1979, greggio1983} that makes the plane of [Mg$/$Fe] vs. Fe a traditionally used diagnostic for the star formation history in the Galaxy. High Mg$/$Fe ratios mark earlier star formation in the halo and thick disc, before [Mg$/$Fe] begin to fall. Since metallicity enrichment runs on similar timescales, populations form the typical knee shape in [Mg$/$Fe] vs. Fe distribution. The downturn position of $\mgfe$ is linked to the star formation intensity and yield  loss rates of star-forming systems. The intense star formation and relatively low mass-loss rates in thick-disc (or central disc in the interpretation of \citealt{schoenrich2009a}) like regimes enable the system to reach high metallicity ($\sim -0.6$ dex) before SN Ia enrichment sets in.  In contrast, the high mass loss/low star formation intensity of dwarf galaxies, which contribute to the halo, sets this knee generally to \feh $< -1$ dex and gives them low $\alpha/$Fe stars at low metallicities \citep[][and references therein]{venn2004, grebel2005, tolstoy2009, matteucci2014}. 

\begin{figure}
\begin{center}
\includegraphics[width=0.5\textwidth, angle=0]{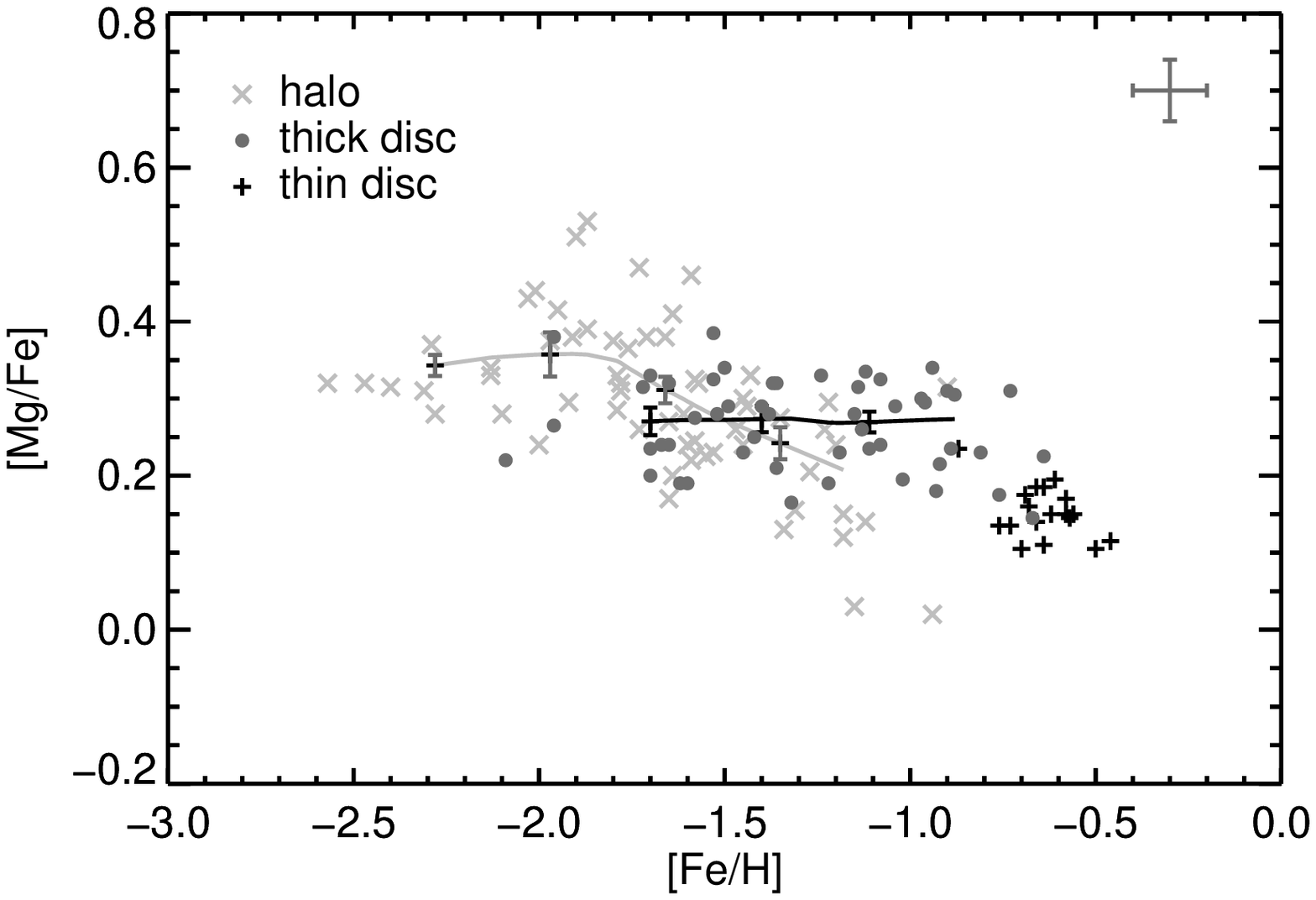}
\caption{$\md$ NLTE [Mg$/$Fe] abundance ratios as a function of metallicity, separated via a kinematic classification from \citet{ruchti2011}. We depict likely halo stars with grey crosses, likely thick disc stars with filled circles, and likely thin disc stars with plus signs. The errors bars represent the sample estimated uncertainty of the mean. The solid lines depicts the smoothed $\mgfe$ distribution computed using LOESS regression.}
\label{pops}
\end{center}
\end{figure}

An interesting question in chemical evolution is if the high-$\alpha$ plateau is really a plateau or shows trends with metallicity. A deviation could be related, for example, to a very fast channel for SN Ia production \citep{matteucci2006} or to a metallicity dependence of SN II yields.  A bimodal distribution of SN Ia delay times, with a significant early contribution from systems with lifetimes of only $10^8$ Gyr, was proposed to explain the dependence of SN Ia rates on galaxy colours \citep{mannucci2005, mannucci2006}, and in the context of the Galactic chemical evolution this scenario implies that $\alpha/$Fe ratio begins to decline already at very low metallicity $\feh \sim -2$. As to SN II, the ratio of \mgi\ and Fe production is very sensitive to the specific SN II piston model \citep{ww1995} and, since Mg stems from shells above the Ni$/$Fe core, critically depends on the cut-off for fall-back of material onto the neutron star/black hole \citep[see][for a discussion of the pre-collapse shell structure]{limongi2003}. Increasing escape of material from the core region of the massive stars (which is linked to the $^{56}$Ni brightness) can hence lower the Mg$/$Fe ratios.

Our NLTE (Fig. \ref{final}, panels c,d) results imply different behaviour of $\alpha/$Fe below \feh $\sim -1$. On the surface, our NLTE determinations show a lower mean [Mg$/$Fe] at low metallicities. For the stars with $-2.5 <$ [Fe$/$H] $< -2$, we get the mean ratio $\mgfe \sim 0.45$ dex in 1D LTE, but only $\mgfe \sim 0.30$ in 1D NLTE or $\md$ NLTE. This happens also in some classical evolution models, e.g. \citet{timmes1995} indicate a depression to a solar value at [Fe$/$H] $< -2$ in all $\alpha$-chain elements. This depression is more prominent if different fall-back schemes and low-energy SN II are taken into account. \citet{brusadin2013} show that two-infall CGE model with the gas outflow during the halo formation phase predicts a depression at $\feh$ from $-1.8$ to $-1.0$ with a "knot"-like structure that indicates the phase in the evolution of the Galactic halo where star formation was inactive. We do not find evidence for a substructure of this kind in our $\md$ NLTE distributions of $\mgfe$. On the other hand, a depleted region in intermediate $\mgfe$ is a natural feature of analytical chemical evolution models \citep{schoenrich2009a,schoenrich2009b}, even more so in radial migration models on the low-metallicity side, where the locally observed high-$\alpha/$Fe ridge is enhanced by outwards migrators from the more populated inner disc.

However, this is not only a change of mean value, but change of the variation of relative Mg abundances in single stars. The star-to-star scatter in $\mgfe$ in the metallicity range $\feh$ from $-2.5$ to $-2$ dex is $0.10$ dex in $\md$ NLTE. However, at $\feh \sim -1.5$ dex, the $\md$ NLTE scatter of $\mgfe$ drops to $0.05$ dex, which is smaller than the internal uncertainties of the measurements, i.e. the measurements are consistent with no cosmic scatter. Also, as already noted earlier, at $\feh \lesssim -1.8$, the dispersion of $\mgfe$ abundance ratios grows. This suggests two options:
i) presence of accreted stars \citep{nissen2010} from low-metallicity dwarf galaxies with extended star formation histories. These are shown to contain numerous low-$\alpha$ stars at low metallicity \citep{shetrone2003,deboer2014,lemasle2014}.
ii) intrinsic scatter, i.e. stochastic chemical evolution, ranging from scatter in local star-formation intensities of galactic protohaloes \citep{gilmore1998} to more sophisticated stochastic chemical evolution models \citep{argast2000, argast2002, karlsson2005a, karlsson2005b, cescutti2008}. Cosmological simulations of the Galaxy formation also predict significant dispersion of $\alpha$-abundance ratios at low metallicity. These results depend on the prescriptions for metal diffusion and metallicity floor, but also with simple prescriptions, $\alpha$-poor and $\alpha$-rich stars have a non-negligible probability to occur at $\feh \sim -2$ and lower \citep{shen2015}.

Finally, in Fig. \ref{pops} we show the $\md$ NLTE derived abundance plane when separated into likely halo vs. thin and thick disc stars via the kinematic classification of \citet{ruchti2011}. To minimize the impact of the systematic uncertainties, we have only plotted those stars, where the 5711 \AA\ line is available; this line, as shown in \citet{bergemann2017} is least sensitive to the effects of convective inhomogeneities and the difference between $\md$ NLTE and full 3D NLTE is within $\sim 0.05$ dex. Thus, the systematic errors are expected to be small. The plot also shows a LOESS curve computed as described in the previous section, but with a larger smoothing parameter, $\alpha  = 0.8$, and ignoring 5 points at the edges, to mitigate the effect of smaller number statistics.
This plot sheds some light on the previously discussed trend of $\mgfe$ vs. $\feh$ at low metallicities: the thick disc ridge does not even show a hint of a slope in $\mgfe$ down to its lowest metallicity members $\feh \approx -1.6$ dex. This makes it difficult to uphold the argument for a fast SNIa channel from low metallicity disc star abundances \citep[][]{matteucci2006}. Instead the trend we saw before in Fig. \ref{final}d is created by the halo populations crossing from on average very high $\mgfe$ and reaching down to near solar abundances around $\feh \approx -1.0$, a behaviour that is observed and described for dwarf galaxies e.g. in \citet{venn2004}, and which is due to their high mass loss rates and very low star formation efficiencies. The increased scatter in the halo $\mgfe$ is consistent with an origin in a diverse population of small accreted galaxies and/or stochastic chemical evolution.
\section{Conclusions}
To shed light on how our understanding of Galactic chemical evolution is impacted by systematics in our assumptions of spectroscopic analysis, we have performed a comparative analysis of [Fe$/$H] vs. [Mg$/$Fe] for a large sample of $326$ halo and thick disc stars using 4 different methods: LTE and NLTE with 1D hydrostatic model atmospheres, as well as with the averages of 3D hydrodynamical model stellar atmospheres. In the $\md$ approach, we take into account the effects of hydrodynamic cooling associated with convective overshooting in the 3D simulations, but the effect of horizontal inhomogeneities is not addressed. However, recent studies \citep[for Mg, see also][]{bergemann2017} suggest that the horizontal fluctuations only have a minor effect on the abundances of elements.

We find that these four modelling scenarios each lead to substantially and qualitatively different trends and distributions of [Mg$/$Fe] with [Fe$/$H]. Compared to $\md$ NLTE results, all other methods lead to significant biases in the $\mgfe$ - $\feh$ plane. The 1D LTE analysis over-estimates the $\mgfe$ abundance ratios at lower metallicities, implying a false mildy increasing trend with declining $\feh$, while the opposite bias happens in 1D NLTE. The differences between 1D NLTE and $\md$ NLTE are also significant at low metallicity, $\feh < -1.5$. This is due to larger differences between the hydrostatic and $\md$ model structures, implying greater (in absolute sense, i.e. more positive or more negative) $\md$ NLTE abundance corrections that was also demonstrated in \citet{bergemann2017}. For red giants, the $\md$ NLTE corrections are $\sim 0.1$ dex compared to 1D NLTE, while for metal-poor dwarfs they change in the opposite direction.

The differences between the trends have profound implications for chemical evolution models (of the Milky Way in this work, and, naturally, for other galaxies), because they imply different formation scenarios for the Galactic components. The declining 1D LTE $\mgfe$ trend, seen already for $\feh \lesssim -2$ may suggests very fast channel for SN Ia production, but this scenario is not supported by our 1D NLTE results, which instead points to a mass- and metallicity-dependent production of Mg in SN II. In our most accurate $\md$ NLTE results, which are are closest to full 3D NLTE,  the halo and the thick disc clearly separate from one another in the $\mgfe - \feh$ plane. Stars of the thick disc occupy a narrow band at a constant mean $\mgfe$ of $\sim 0.3$ dex. The thick disc extends to $\feh \approx -1.6$, showing no trend of $\mgfe$ with metallicity and an intrinsic dispersion of less than $0.03$ dex, which argues againt the fast SN Ia channel. In contrast, the halo population shows a trend with $\feh$, with a significant number of very metal-poor stars with high $\mgfe$, but reaching to solar $\mgfe$ values around $\feh \approx -1.0$. This behaviour is observed in dwarf galaxies being due to the their high mass loss rates and very low star formation efficiencies.

Our results are still based on a small sample of stars and that we use mean 3D hydrodynamical models. In this paper we employ only those \mgi\ lines that are least sensitive to the model physics, in particular the 5711 \AA\ line, for which the 3D - $\md$ abundance differences are expected to be within $0.05$ dex. Unfortunately, the few very metal-poor stars in our sample, which have very low (solar-like) $\md$ NLTE $\mgfe$ ratios are also those, where only the 5528 \AA\ line could be measured in the spectra. \citet{bergemann2017} showed that for this spectral line the abundances based derived using full 3D NLTE calculations can be higher by up to $+0.2$ dex. Thus we cannot currently conclude on whether there are truly solar-like $\alpha$-poor stars at very low $\feh < -2$ in the halo. Also, our sample was pre-selected against metal-rich stars, and there are only a few members of the thin disc that does not allow us to probe the thin-thick disc transition. 

Our study has demonstrated how vital full spectroscopic analysis is for the understanding of trends of abundance ratios with metallicity in different stellar components. The systematic biases inferred from simplifying assumptions (1D, LTE) are strong enough to force wrong formation scenarios in chemical evolution models. This paper also shows the urgent need to analyse all important elements in full 3D NLTE models, and to compare their results with Galaxy formation and chemical evolution models to test the validity of their assumptions and predictions.
\acknowledgements
We thank T. Gehren, and F. Grupp for providing the observed high-resolution data and stellar atmosphere models used in this work. RC acknowledges partial support from a DECRA grant from the Australian Research Council (project DE120102940). Funding for the Stellar Astrophysics Centre is provided by The Danish National Research Foundation. GRR acknowledges support from the project grant "The New Milky Way" from the Knut and Alice Wallenberg Foundation. MB acknowledges support by the Collaborative Research center SFB 881 (Heidelberg University) of the Deutsche Forschungsgemeinschaft (DFG, German Research Foundation). This work was supported by the research grants (VKR023406, VKR023371) from Villum Fonden and from Augustinus Fonden. This work has made use of data from the European Space Agency (ESA) mission {\it Gaia} (\url{http://www.cosmos.esa.int/gaia}), processed by the {\it Gaia} Data Processing and Analysis Consortium (DPAC, \url{http://www.cosmos.esa.int/web/gaia/dpac/consortium}). Funding for the DPAC has been provided by national institutions, in particular the institutions participating in the {\it Gaia} Multilateral Agreement. MB thanks Brad Gibson and Gareth Few for useful discussions about the chemical evolution of the Galaxy and for  pointing out the low-energy supernova scenario. We used the IDL software packages from the Coyote graphics library, IDL Astronomy User's Library, and \url{http://www-astro.physics.ox.ac.uk/~mxc/software/}.


\clearpage

\section{Appendix}
\pagestyle{empty}
\begin{figure*}[!ht]
\hbox{
\includegraphics[width=0.35\textwidth, angle=-90]{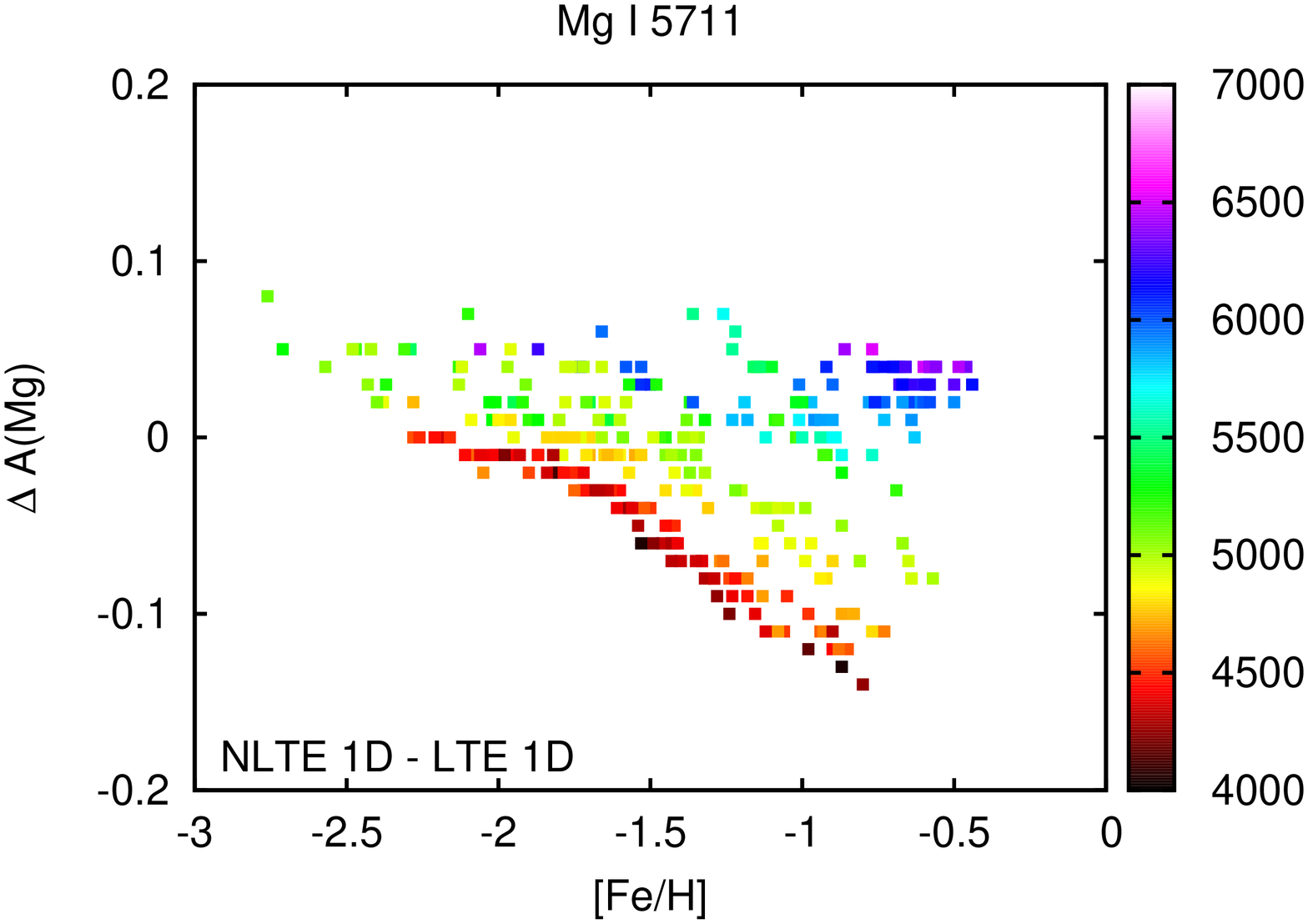}
\includegraphics[width=0.35\textwidth, angle=-90]{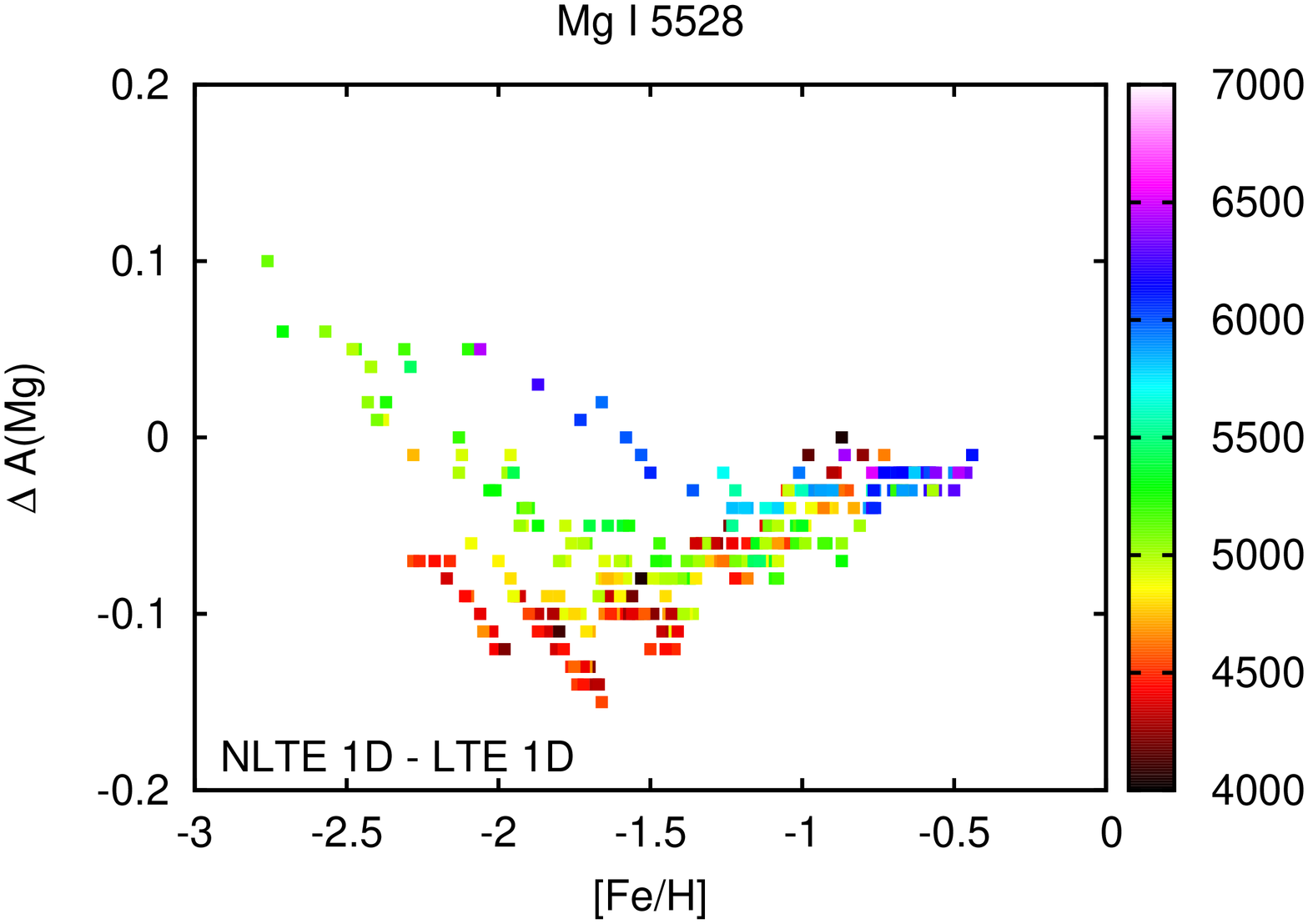}}
\hbox{
\includegraphics[width=0.35\textwidth, angle=-90]{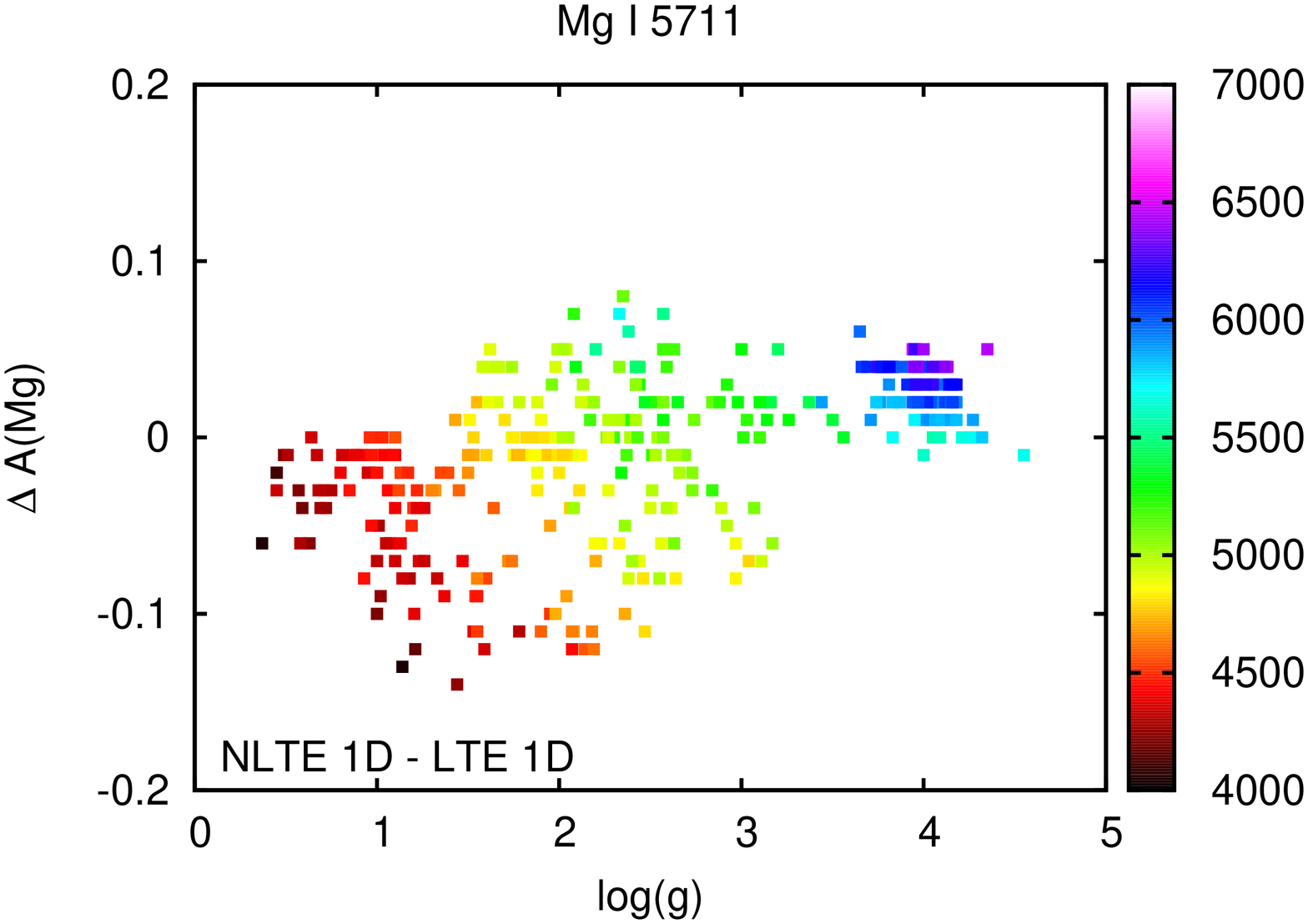}
\includegraphics[width=0.35\textwidth, angle=-90]{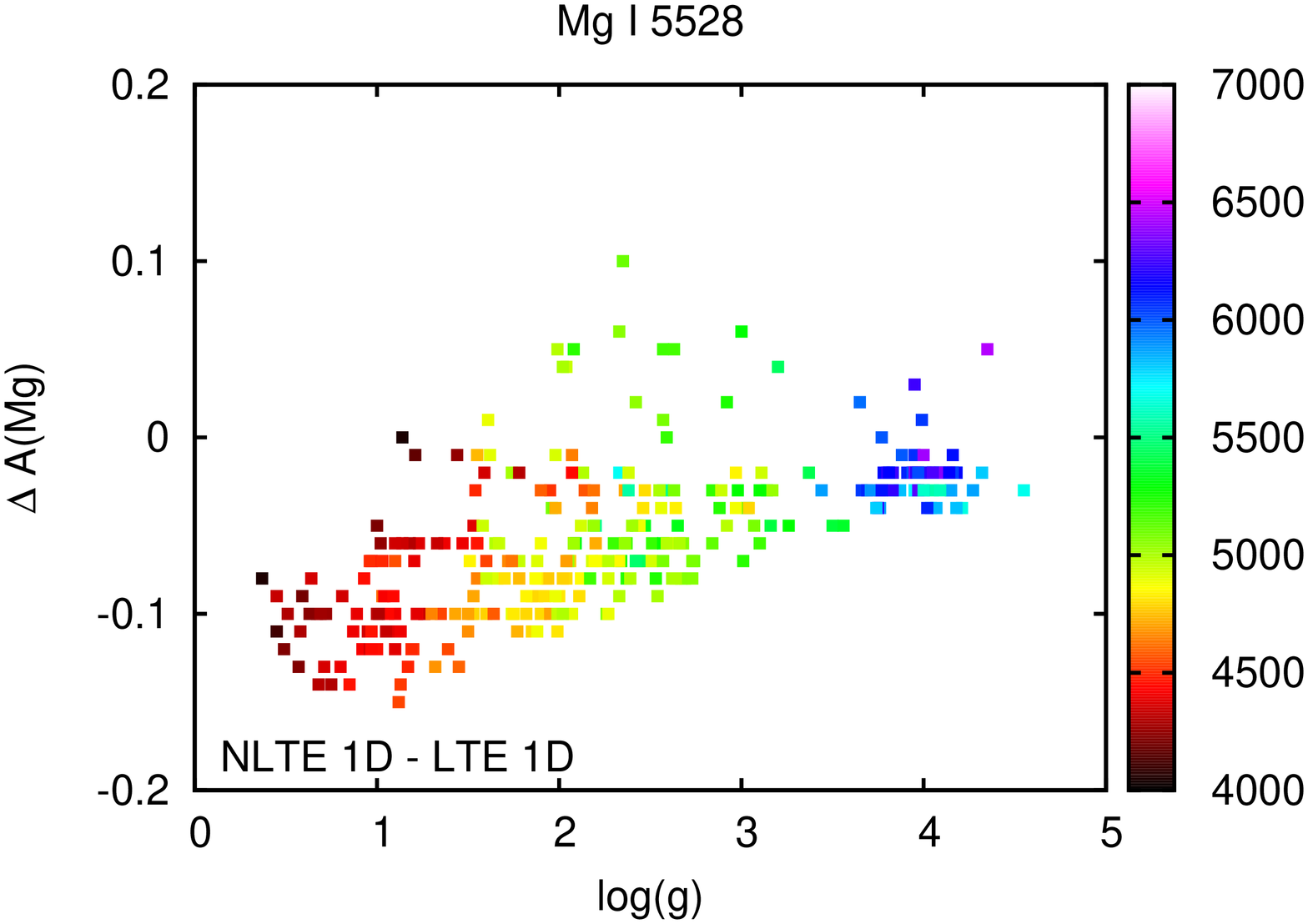}}
\hbox{                                               
\includegraphics[width=0.35\textwidth, angle=-90]{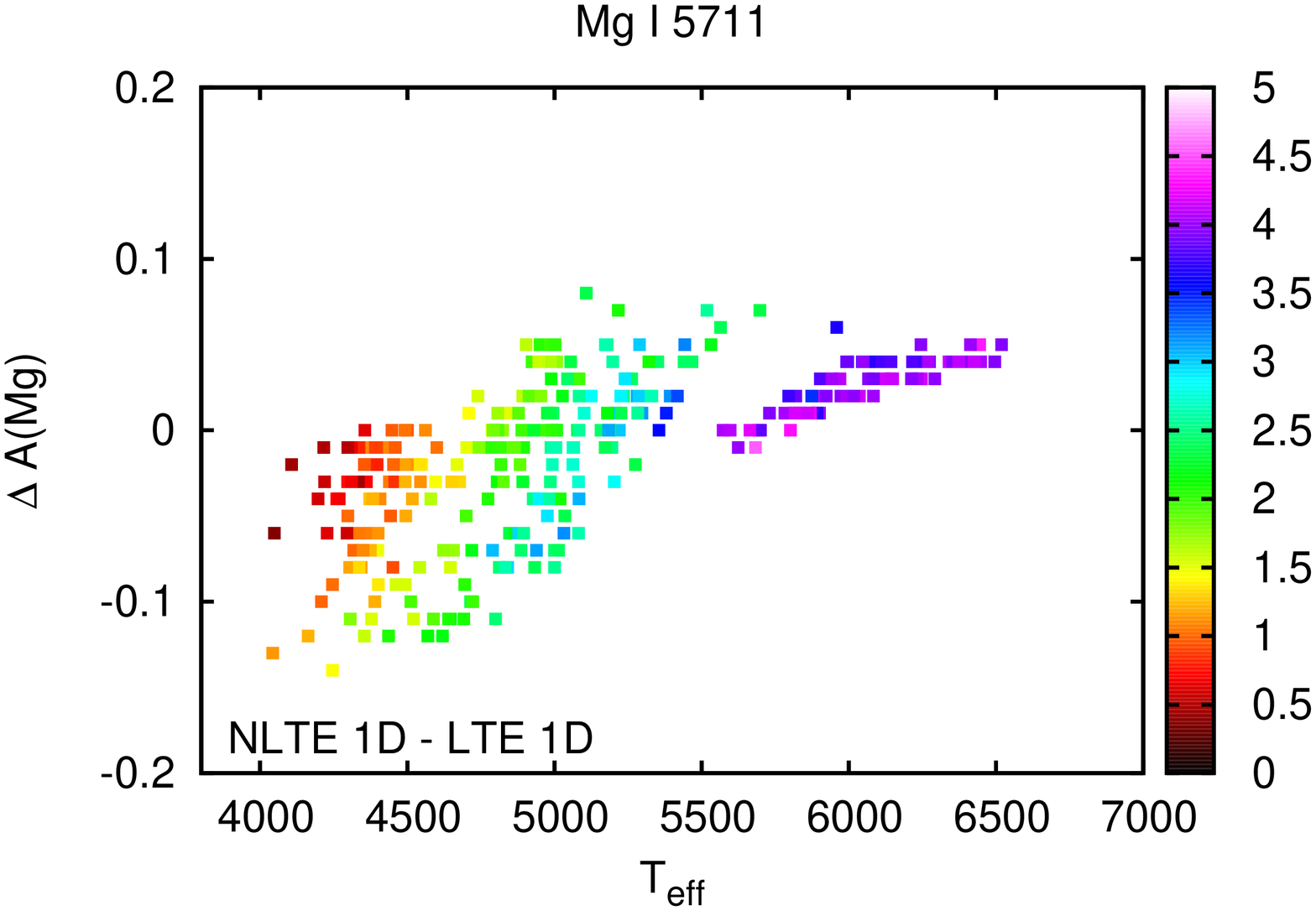}
\includegraphics[width=0.35\textwidth, angle=-90]{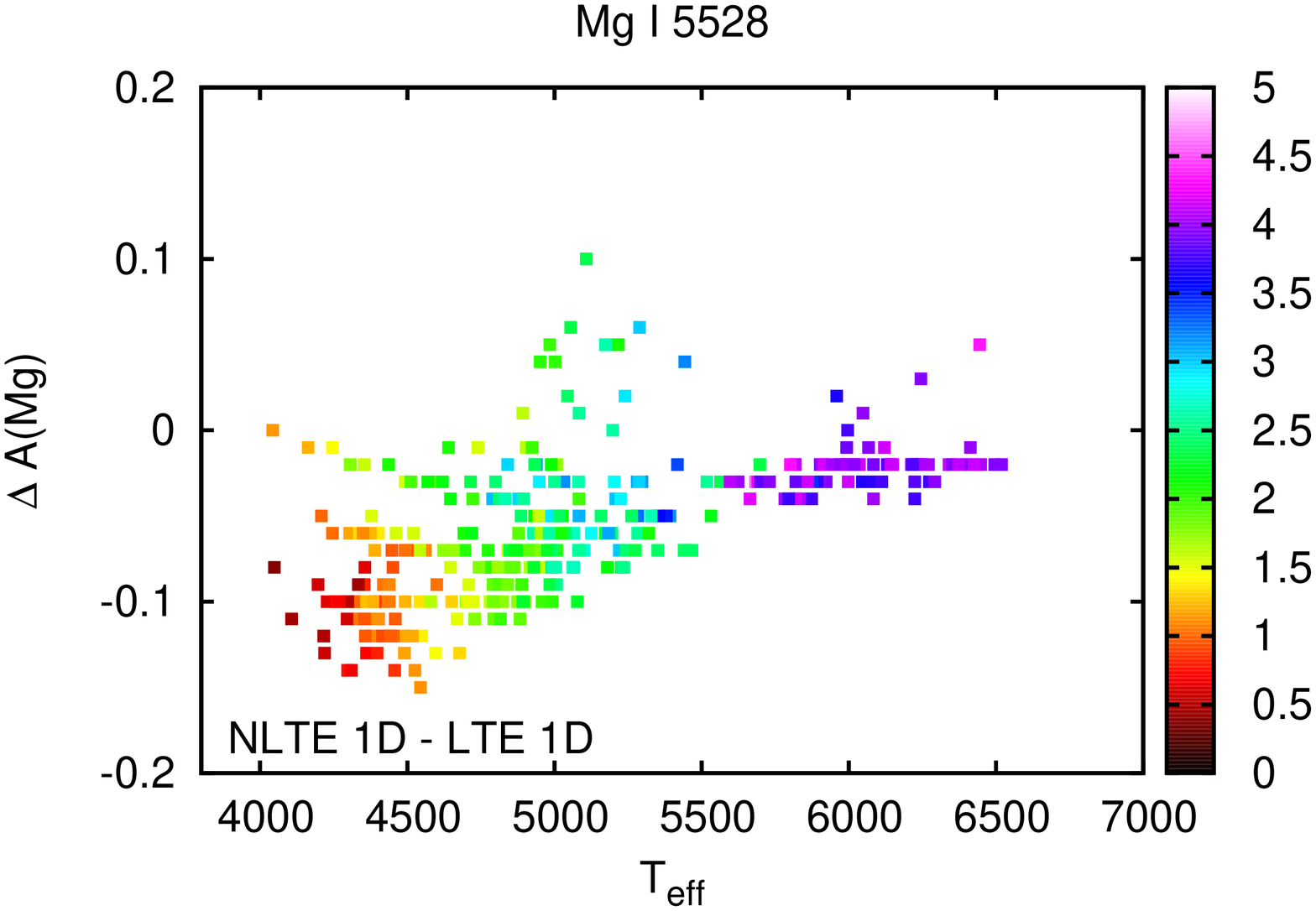}}
\caption{NLTE abundance corrections for the selected \mgi\ lines computed with 1D hydrostatic model atmospheres. The stars are colour-coded with their values of $\teff$ (top and middle panels) and $\log g$ (bottom panels).}
\label{abcor1}
\end{figure*}
\begin{figure*}[!ht]
\hbox{
\includegraphics[width=0.35\textwidth, angle=-90]{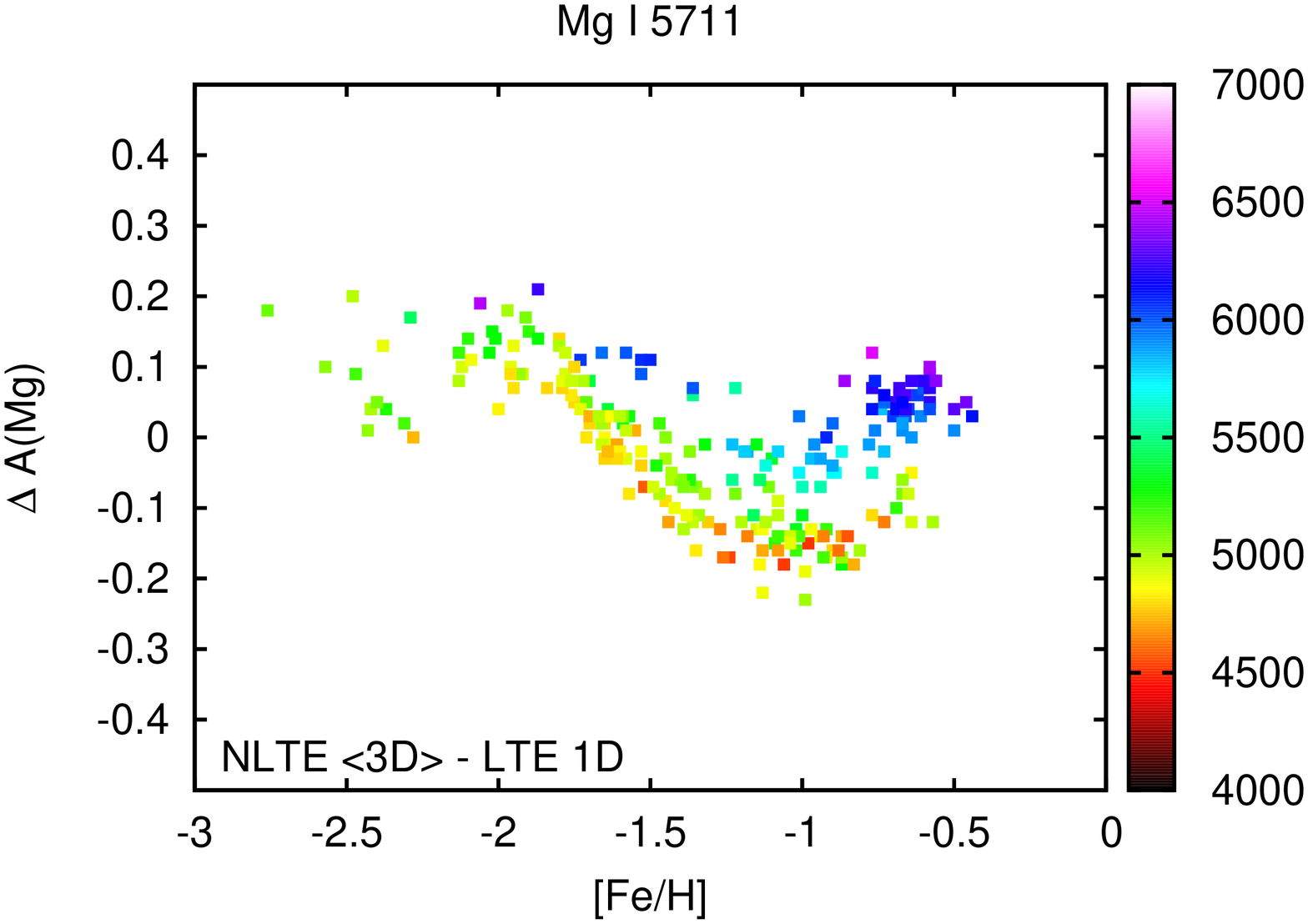}
\includegraphics[width=0.35\textwidth, angle=-90]{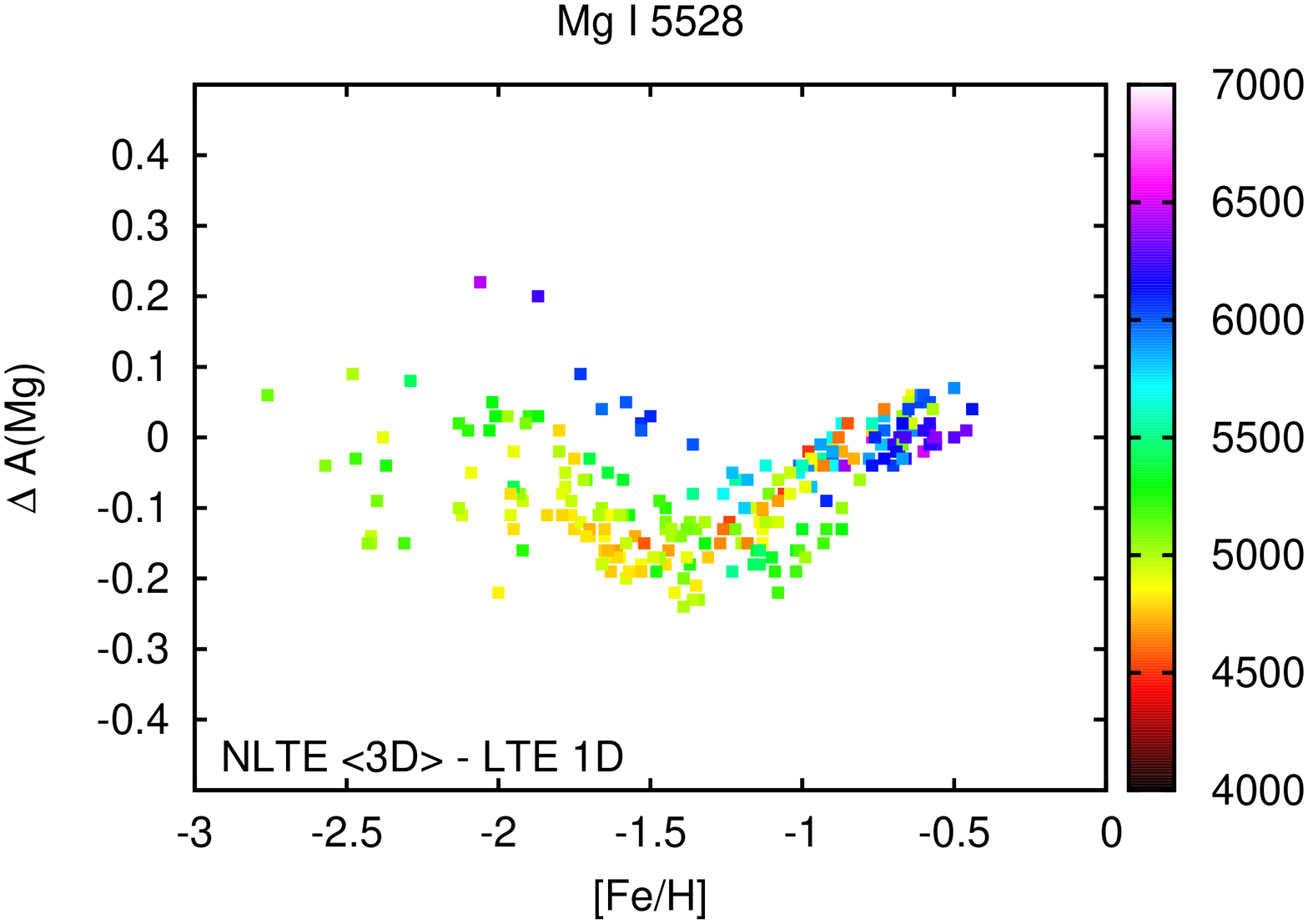}}
\hbox{
\includegraphics[width=0.35\textwidth, angle=-90]{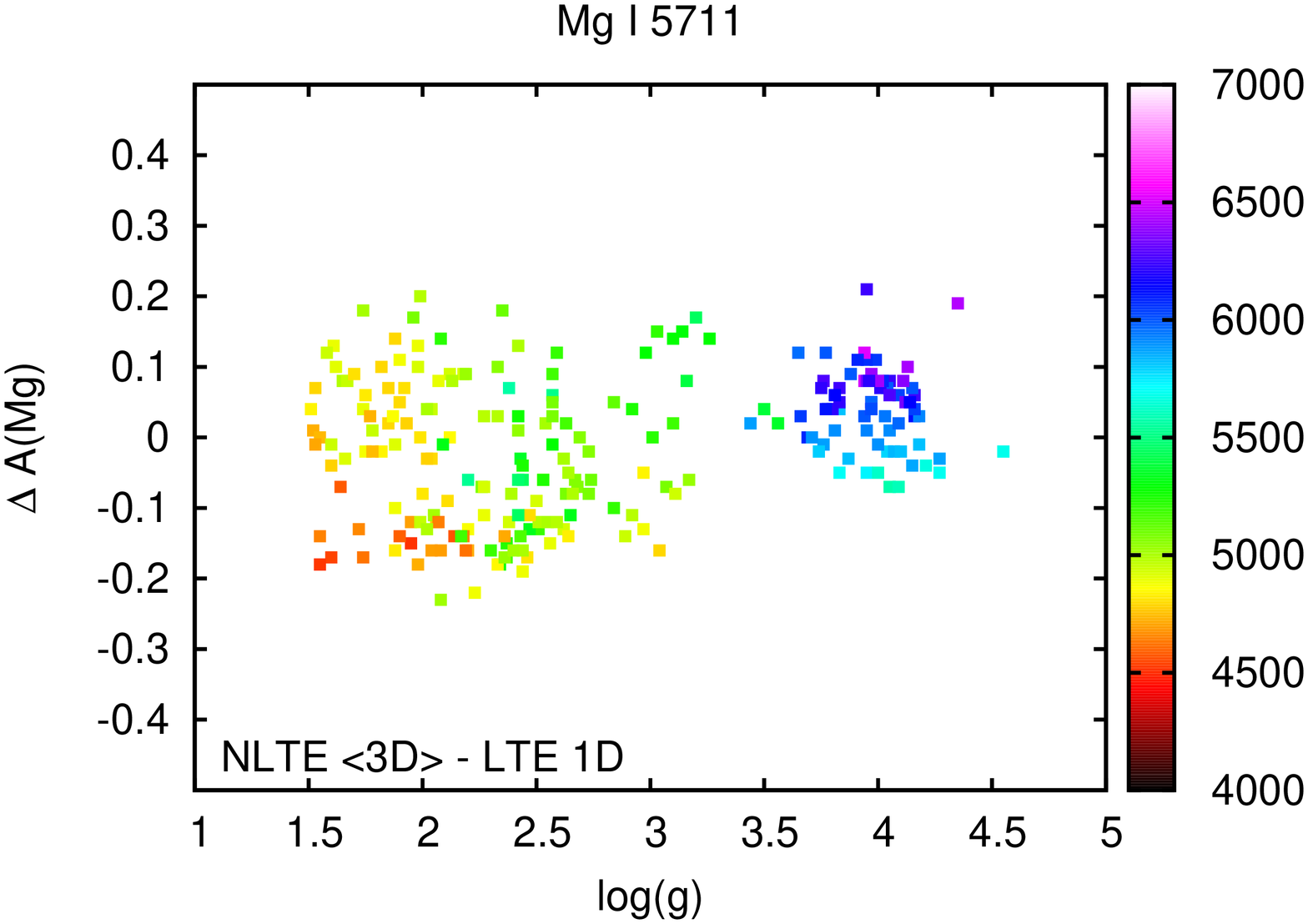}
\includegraphics[width=0.35\textwidth, angle=-90]{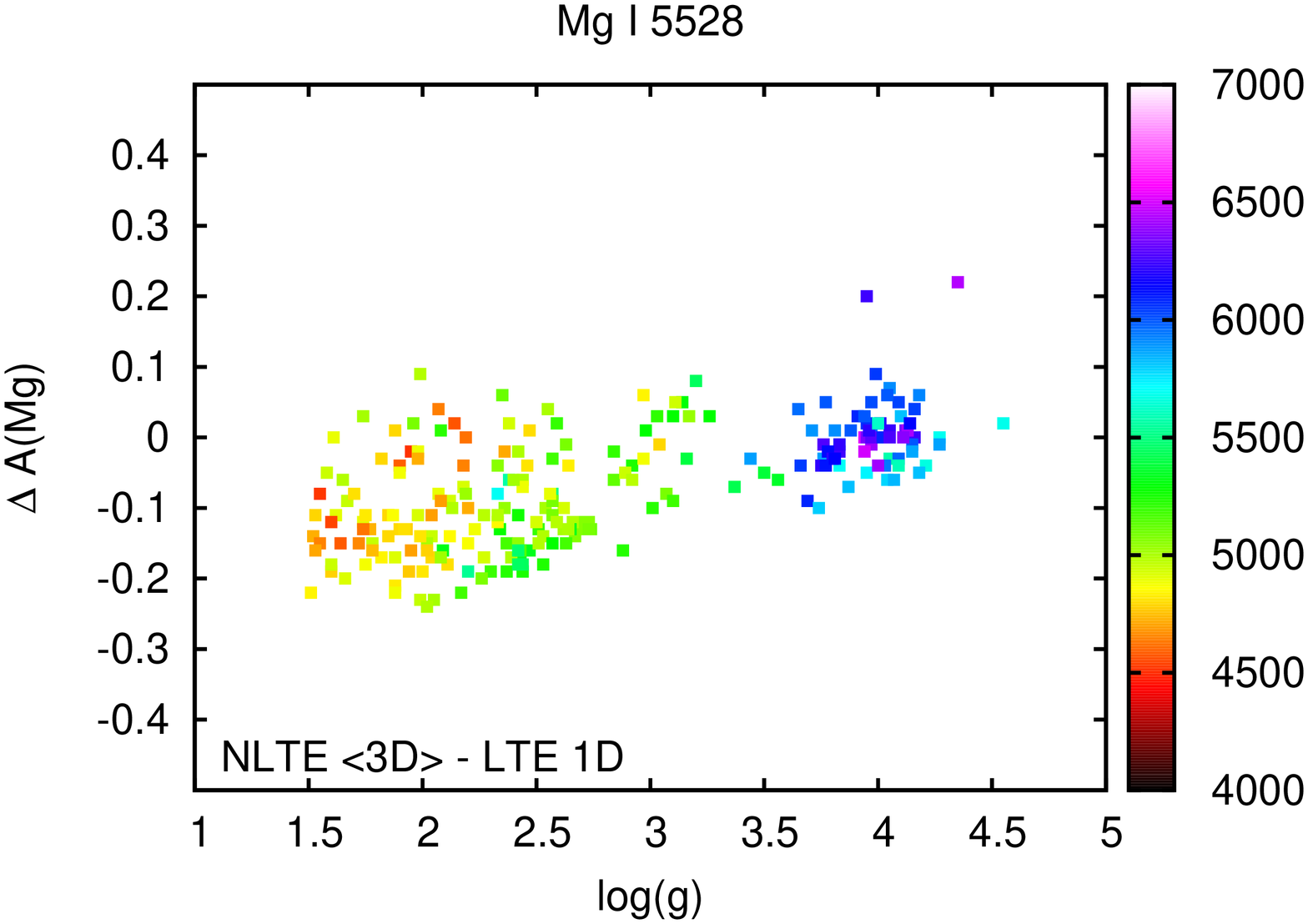}}
\hbox{                                               
\includegraphics[width=0.35\textwidth, angle=-90]{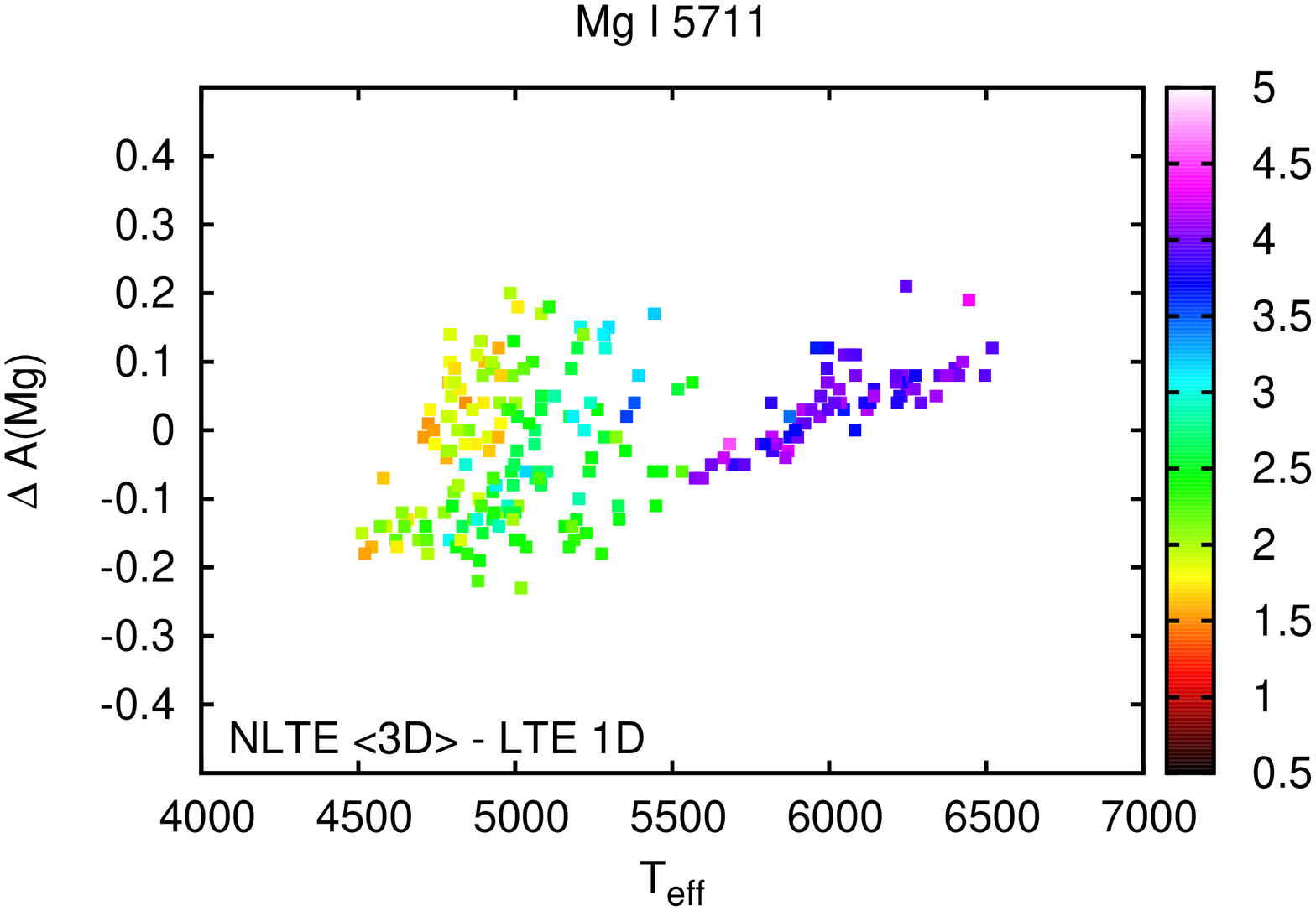}
\includegraphics[width=0.35\textwidth, angle=-90]{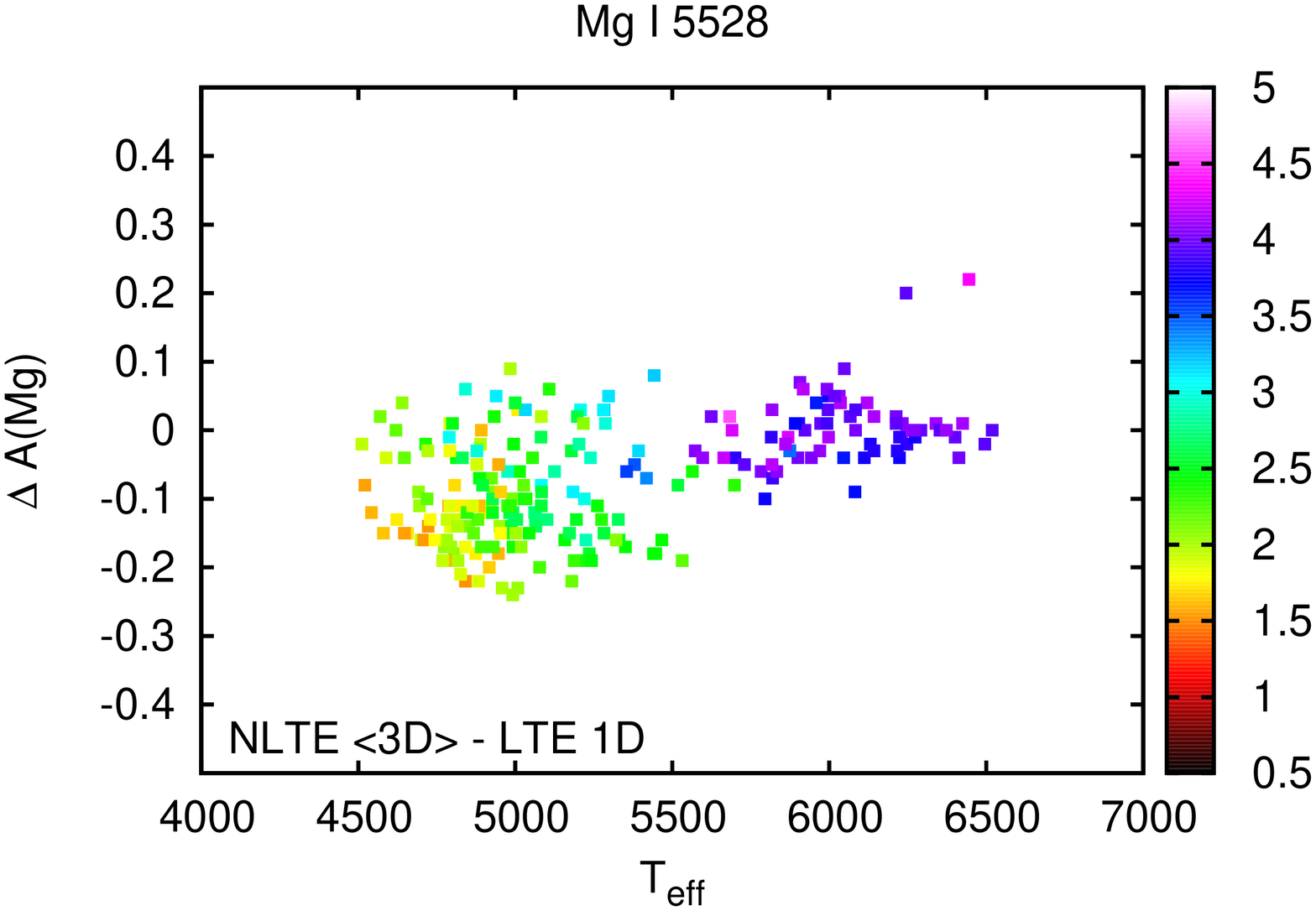}}
\caption{NLTE abundance corrections for the selected \mgi\ lines computed with $\md$ model atmospheres. The stars are colour-coded with their values of $\teff$ (top and middle panels) and $\log g$ (bottom panels).}
\label{abcor2}
\end{figure*}
\end{document}